\documentclass[12pt,preprint]{aastex}
\usepackage[usenames]{color}
\usepackage{rotating}

\def\l{$\lambda$}
\def\mbh{$M_{\rm BH}$\/}
\def\nh{$n_{\mathrm{H}}$\/}
\def\lledd{$L/L_{\rm Edd}$}

\def\nc{$N_{\rm c}$\/}
\def\rfe{$R_{\rm FeII}$}

\def\feiiq{\rm Fe{\sc ii}$\lambda$4570\/}

\def\ltsima{$\; \buildrel < \over \sim \;$}
\def\ltsim{\lower.5ex\hbox{\ltsima}}  
\def\gtsima{$\; \buildrel > \over \sim \;$}

\def\gtsim{\lower.5ex\hbox{\gtsima}}

\def\civ{{\sc{Civ}}$\lambda$1549\/}

\def\cmq{cm$^{-2}$\/}
\def\cm3{cm$^{-3}$\/}
\def\hb{{\sc{H}}$\beta$\/}
\def\hg{{\sc{H}}$\gamma$\/}
\def\hbbc{{\sc{H}}$\beta_{\rm BC}$\/}

\def\hbnc{{\sc{H}}$\beta_{\rm NC}$\/}
\def\mgii{{Mg\sc{ii}}$\lambda$2800\/}

\def\ciii{{\sc{Ciii]}}$\lambda$1909\/}
\def\oiiiopt{{\sc{[Oiii]}}$\lambda\lambda$4959,5007\/}
\def\o4363{{\sc{[Oiii]}}$\lambda$4363\/}
\def\oii{{\sc{[Oii]}}$\lambda$3727\/}

\def\fei{{Fe{\sc i}}}

\def\siiii{Si{\sc iii]}$\lambda$1892\/}
\def\aliii{Al{\sc iii}$\lambda$1860\/}

\def\feiiuv{{{\sc{Feii}}}$_{\rm UV}$\/}
\def\feiiopt{{Fe \sc{ii}}$_{\rm opt}$\/}
\def\feii{{Fe\sc{ii}}\/}

\def\fe{{\sc{Fe}}\/}

\def\vr{{$v_{\mathrm r}$}}

\def\fe76087{{\sc [Fe vii]}$\lambda$6087\/}
\def\oiii{{\sc [Oiii]}$\lambda$5007}

\def\kms{km~s$^{-1}$}

\def\ergss{ergs s$^{-1}$\/}

\def\heii{{{\sc H}e{\sc ii}}$\lambda$4686\/}

\def\siiv{Si{\sc iv}$\lambda$1397\/}
\def\oiv{O{\sc iv]}$\lambda$1402\/}

\slugcomment{}

\shorttitle{\mgii}
\shortauthors{}

\begin{document}
\title{Low-Ionization Outflows in  High Eddington Ratio Quasars}

\author{Paola Marziani\altaffilmark{1}, Jack W. Sulentic, Ilse Plauchu-Frayn, Ascensi\'on del Olmo} %\altaffilmark{1}}
\affil{Instituto de Astrof\'isica de Andaluc\'ia, CSIC, Granada, Spain}
\altaffiltext{1}{INAF, Astronomical Observatory of Padova, Padova, Italy. e-mail paola.marziani@oapd.inaf.it}

\begin{abstract}
The broad \mgii\ doublet has been much studied in connection with its
potentially
important role as a virial estimator of black hole mass in high redshift
quasars. An important task is therefore identification of any line components
likely related to broadening by non-virial motions. High s/n median composite
spectra (binned in the ``4D eigenvector 1'' context of \citealt{sulenticetal07}) were constructed for the
brightest 680
SDSS DR7 quasars in the $0.4 \le z \le 0.75$\ range where both \mgii\ and
\hb\ are
recorded in the same spectra.  Composite spectra representing 90\%\ of the
quasars confirm
previous findings that FWHM(\mgii) is about 20\%\ narrower than FWHM(\hb).
The situation is clearly different for the most extreme (Population A)
sources which
are the highest Eddington radiators in the sample. In the median spectra of these sources FWHM
\mgii\ is
equal to or greater than FWHM(\hb) and shows a significant blueshift 
relative to \hb.  We interpret the \mgii\ blueshift as the signature of a
radiation-driven wind or outflow  in the highest accreting quasars. In
this interpretation { the \mgii\ line width -- affected by blueshifted
emission -- is unsuitable for virial mass estimation  in $\approx$ 10\%\
of quasars.}

%All 4DE1 parameters are known to be extreme in these bins with  \civ\ showing larger blueshifts/asymmetries and stronger soft X-ray excesses.

% and  high s/n composite spectra can be computed. 
%The resultant composite spectra yield essentiallly rms  measures for a restricted range of source luminosity  ($<$log L$_{BOL}>$=46.0) covered by our sample.The general details  of our H$\beta$ and MgII comparison are given elsewhere.

\end{abstract}


\section{Introduction}
\label{intro}

Estimation of black hole mass (\mbh) and Eddington ratio (\lledd) for quasars is of great interest both to researchers working on models of the broad line region (BLR) structure and to cosmologists. Therefore we need a large number of accurate as possible estimates over the widest possible range of redshift, source luminosity and line/continuum properties. FWHM(\hb) is the principal virial estimator at low redshifts. While \hb\  can be followed into the infrared (out to $z \approx 3.7$) at least one additional line is needed to provide  complementary \mbh\ estimates over the full redshift range where quasars are observed. \mgii\ is the best candidate since it is a low-ionization line and avoids some of the difficulties associated with \civ\ \citep[e.g.][]{netzeretal07,sulenticetal07,marzianisulentic12,trakhtenbrotnetzer12,denney12}.

The advent of the SDSS database makes possible a direct calibration of \mgii\ using sources where {\em both} \mgii\ and \hb\ appear in the same SDSS spectra.  Only high s/n composite spectra can  provide an ideal vehicle for such comparisons. In addition one has to consider that source diversity found within the context of a formalism like  4D Eigenvector 1 \citep[4DE1][]{sulenticetal00a,sulenticetal07, marzianietal10} is large and likely driven  by Eddington ratio \lledd \citep[e.g.][]{borosongreen92,marzianietal01,boroson02,grupe04,kruzceketal11,tangetal12}. { The 4DE1 allows to discriminate sources whole emission line profiles and whose spectrophotometric properties are strikingly different. A notable empirically-motivated boundary at low- and moderate luminosity is set by  FWHM(\hb) $\approx$4000 \kms. Sources narrower than this limit show \hb\ profiles that are well fit by a Lorentzian function, while broader sources show prominent redward asymmetries in their \hb\ profiles \citep[e.g.][]{veroncettyetal01,marzianietal03b}. Sources with FWHM(\hb) $\le$ 4000 \kms\  include   Narrow Line Seyfert 1s (NLSy1s) by definition and  are characterized by a significant high-ionization outflow, revealed by a   \civ\ blueshift with respect to rest frame or to broad low-ionization lines \citep[e.g.][]{gaskell82,tytlerfan92,marzianietal96,sulenticetal07,richardsetal11}.  The outflow  has been ascribed to a radiative or magnetically  driven wind \citep[e.g.][]{murraychiang97,bottorffetal97,progaetal00}. }  Therefore it is important to at least distinguish  between sources that appear to be wind- or disk-dominated \citep{richardsetal11} applying the  limit at FWHM(\hb) = 4000 \kms\ that separates Population A and B sources (\citealt{sulenticetal00a}, c.f. \citealt{collinetal06}). 

{ A finer subdivision is still needed even with the restriction to Population A.} Pop. A sources  span a relatively large range in  Eddington ratio, $\Delta \log$\lledd$\approx$0.5, that likely involves not only  the highest Eddington radiators.  We apply the spectral classification of \citet{sulenticetal02} that divides the plane FWHM(\hb)--\rfe\ into bins of $\Delta$\rfe = 0.5 and $\Delta$FWHM(\hb) = 4000 \kms, where \rfe\ is computed as the equivalent width (or intensity) ratio of the \feiiq\  blend and broad \hb\ (see their Fig. 1).   Extreme Pop. A (A3 and A4) sources with \rfe$\gtsim$1  are characterized by the strongest high-ionization outflow, with the largest \civ\ blueshifts \citep{marzianietal06,sulenticetal07}, and are believed to be the highest Eddington ratio sources.  A large low-$z$\ sample covering the \hb\ and \mgii\ emission lines is defined (\S \ref{sample}). We point out intriguing changes of the \hb\ and \mgii\ line profile occurring in bins A3 and A4 (\S \ref{results}) and discuss first order considerations about the physics involved (\S \ref{discussion}).

\section{Sample Selection and Construction of Composite Spectra}
\label{sample}

We searched SDSS-DR7  for sources catalogued as Type 1 AGN (quasars)  in the redshift range  0.4 -- 0.75 and with magnitudes brighter than $g \approx$ 18.5 in the $g$, $r$ or $i$\ bands, as well as the \citet{zhouetal06} catalog. The resultant sample  consisted of 716 quasars reduced to 680 (all of Pop. B: 369 sources; A1: 97, A2: 156, A3: 43, A4:15) by discarding very noisy spectra and some sources with unusually red colors.  Broad Absorption Line (BAL) QSOs were excluded from the sample.   
%Within the same $z$\ and $m$\ limits we added 16 sources from the  sample that were misclassified in the SDSS catalog. 
%Below this magnitude limit the low s/n of individual spectra make it impossible to estimate FWHM \hb\ and \rfe\ well enough to assign  them to one of our pre-defined bins. 
%Discarding very noisy spectra and some sources with unusually red colors (obviously reddened quasars where \mgii\ was very weak)  reduced  the sample to 680 quasars. 

The rest frame was set by measuring the wavelengths of  three of the most prominent narrow lines (\oii, \hb, and \oiii) when they were detected.  Residual systematic wavelength shifts \citep{hewettwild10} in addition to the SDSS-provided redshift values were computed taking an  average of the three lines in each source spectrum,  clipping  individual measurements in cases of disagreement because of poor data or intrinsic blueshift of \oiii\ \citep{huetal08a}.   We used IRAF {\sc splot} to estimate FWHM(\hb) in order to better separate sources into spectral bins following the prescription in \citet{sulenticetal02}.  Assignments for all bins   were made by visual inspection of each spectrum and estimation of  \rfe\ through {\sc ngaussfit}. Median composites were constructed  respectively for H$\beta$ and \mgii\ after redshift correction and continuum normalization at 5050 \AA\ and 3050 \AA.  
%Combining sources of uncertainty (SDSS wavelength calibration, limited s/n in the narrow lines) 
The  rest frame radial velocity of the composites (defined by the average of peak radial velocity of narrow \oii, H$\delta$, \hg, \hb, and \oiiiopt) was found  $<$ 10 \kms\ in A1 and A2, and $<$ 20 \kms\ in bin A3 and A4, with a rms value always less than 50 \kms. A line can be considered unshifted with respect to rest frame if $|\Delta$\vr$|$  $\le 100$ \kms.   { The relative uncertainty of \vr\ measurements for \hb\ and \mgii\ peak
velocities on the composite spectra is somewhat less and has been estimated by
propagating 3 sources of error on: (1) zero-point;  \oii\ and \hb\ narrow
component peak wavelengths measured in
the four composites agree within rms $\pm$10 \kms; (2) wavelength
calibration, by measuring the dispersion in wavelength measurements of
strong sky lines, typically $\pm 15$ \kms. Systematic shifts are
consistent with 0 ($\ltsim$ 25 \kms) for HgI \l 4359 and OH \l 8401 lines whose wavelengths correspond roughly to the wavelengths of the redshifted \mgii\ doublet and \hb\ line, respectively;
(3)  peak line position determined by the multicomponent fit, as provided by the fitting
program {\sc specfit} (\S \ref{results}).
 }

\begin{figure}
\includegraphics[scale=0.35]{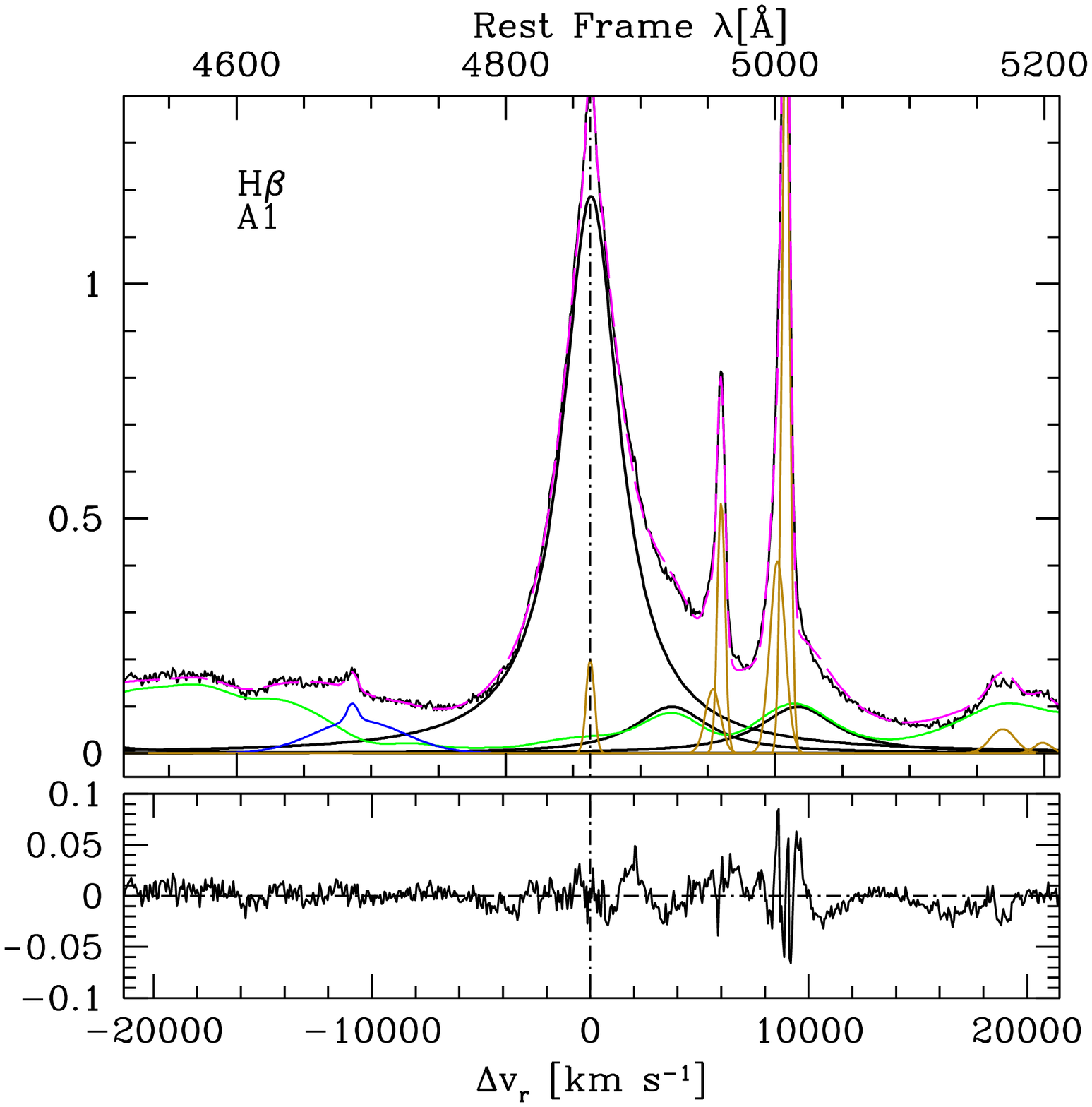}
\includegraphics[scale=0.35]{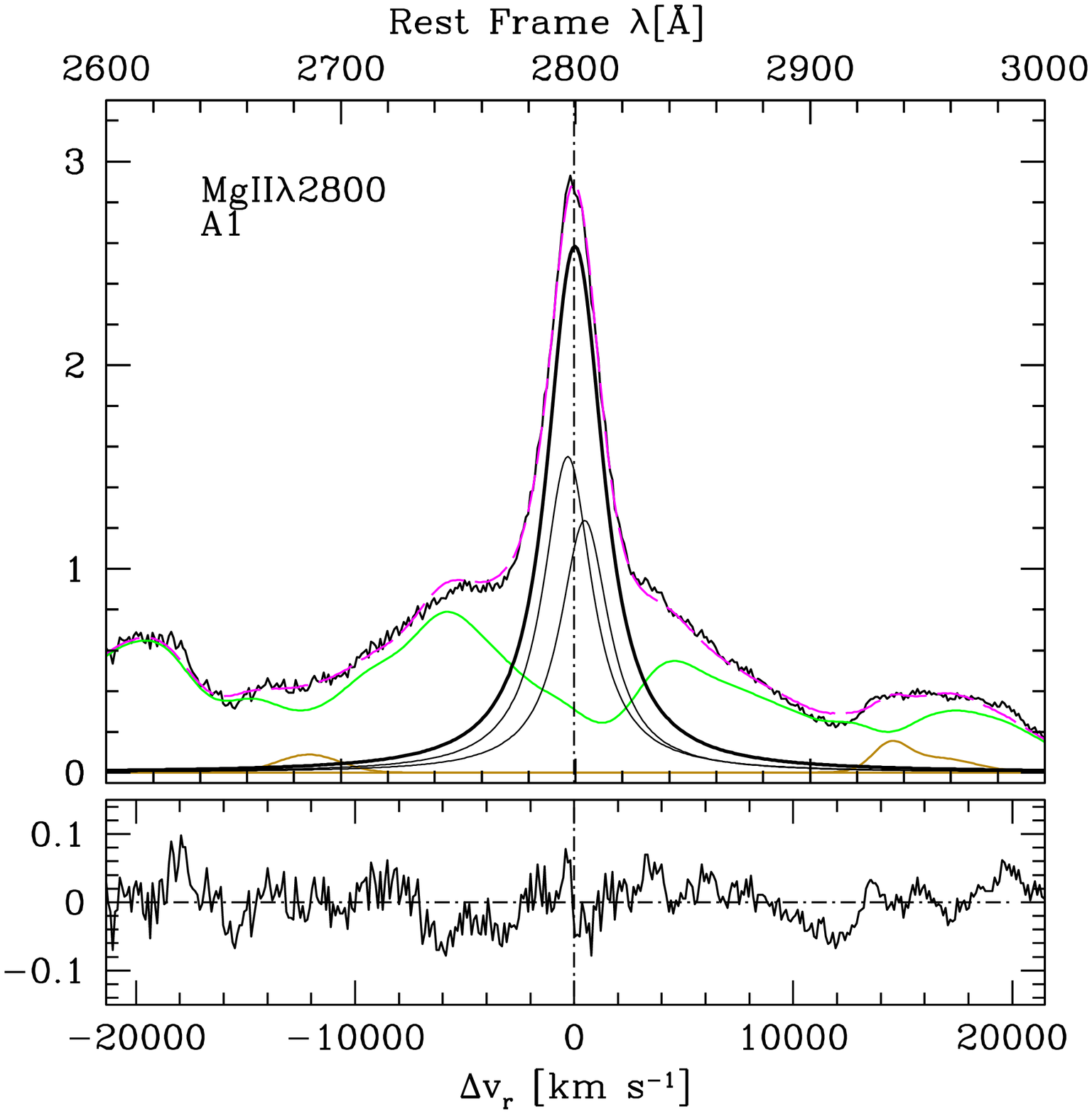}\\
\includegraphics[scale=0.35]{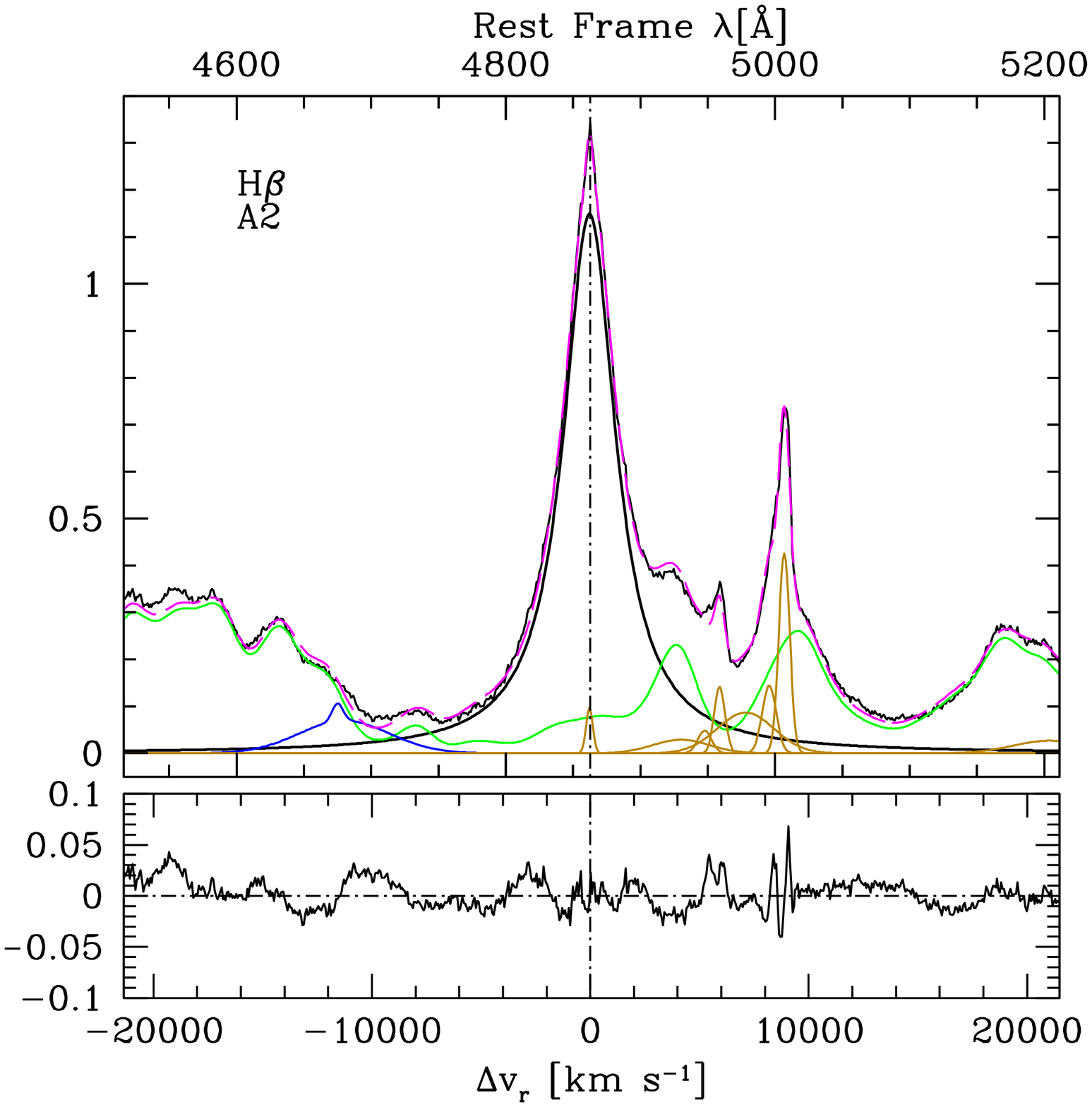}
\includegraphics[scale=0.35]{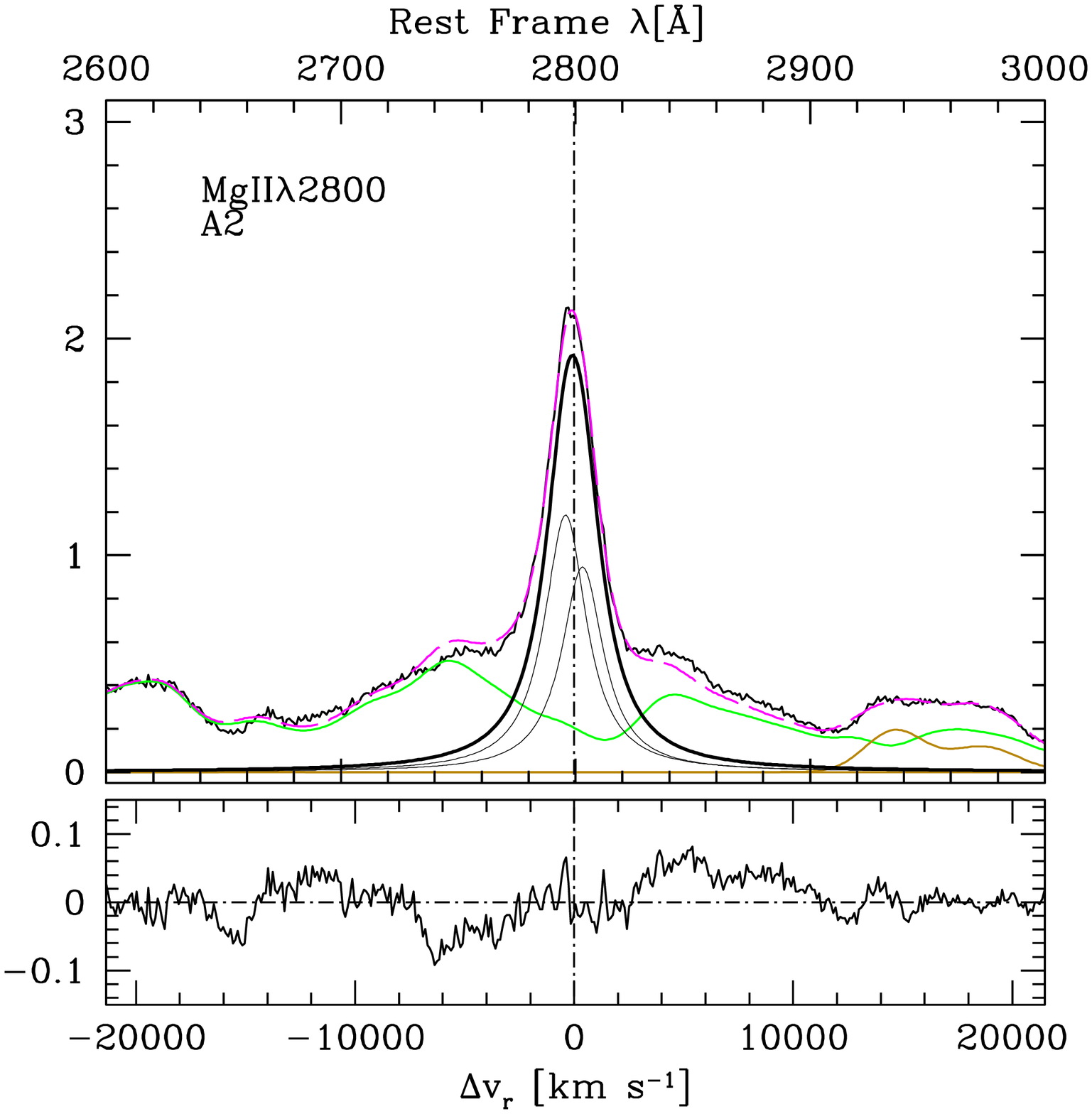}\\
\caption{{ Spectra of \hb\ (left panels) and \mgii\ (right panels) for
spectral types A1 (top) and A2 (bottom).
The horizontal scale is rest frame wavelength [\AA] or radial velocity
with the origin indicating rest frame
(laboratory) wavelength.  In the \mgii\ panels the vertical  dot-dashed
line is drawn at the reference
wavelength 2799.4 which corresponds to a component ratio 1.25:1.00. The
black lines show the original
continuum-subtracted spectrum while the dashed magenta line shows the
model including all emission line components.
The thick black lines show the broad component and thin black lines the
individual components of the \mgii\ doublet.
The green lines trace \feiiopt\ and \feiiuv\ emission and the gold-brown
lines various contributions associated
with the narrow-line region  (\hbnc, \oiiiopt). In the \hb\ panels significant \heii\ is revealed (thick blue line).  In the A1 \hb\ panel  \ion{He}{1}\l 4924 and \ion{He}{1}\l 5016 (black lines) almost overlap with the m42 \feiiopt\ lines.  The \fei\ + \ion{He}{1}\l 2945  emission (modeled as the sum of two Gaussians) is also traced by a brown line and is visible toward the right end of the \mgii\ panel at $\approx$ 2950\AA.}}
\label{fig:hbmga1a2}
\end{figure}

{ Interpretation of the \hb\ spectral range closely follows previous work
\citep{borosongreen92,marzianietal03a,marzianietal09}. Along with continuum and \feii\ emission fit over the spectral range 4430 -- 5510 \AA,
we will include possible contribution of \ion{He}{1} lines at 4471 ad 5016 \AA\ (that appear to be significant only for bin A1), \heii\ and a contribution  due to [\ion{Fe}{7}] and [\ion{N}{1}] lines at $\approx$ 5150 -- 5200 \AA. The region around \mgii\ has been studied by several authors since the mid-1980s
\citep{wampler85,brothertonetal94,grahametal96,laoretal97a,vestergaardwilkes01}.
Emission blends near \mgii\ are mainly due to \feii. We define a range for
{\sc specfit} analysis (2600--3050 \AA) that is a compromise between
proximity to the line and
the necessity of having sufficient continuum coverage to properly map the
broad FeII blends. We also include other known lines e.g. semi-forbidden
\ion{Al}{2}] 2669.95
and { \ion{O}{3} 2672.04.} in the fits. \citet{bruhweilerverner08} provide
\feiiuv\ emission
templates computed from  {\sc cloudy} simulations
and using an 830 level model of the Fe$^{+}$ ion. Use of the
\citet{bruhweilerverner08} \feiiuv\ template results in a
systematic residual near 2950 \AA\ (Figs. \ref{fig:hbmga1a2} and
\ref{fig:hbmga3a4}). The
excess flux is probably due to the blend of \ion{He}{1}\l 2945 and \fei\ emission from a cluster of lines
produced by transitions from  the terms z$^{5}$F and a$^{5}D^\mathrm{o}$
to the ground state (a$^{5}D$).   \fei\ emission  has been
predicted by photoionization models \citep{sigutetal04}
and was suggested by previous observations \citep[e.g.,][]{kwanetal95,grahametal96}. The flux deficit is larger  than the \fei\ predicted by photoionization models, at least by a factor of several with respect to a low-ionization, 3 times solar metallicity case \citep[model {\sc u20h11} of ][]{sigutetal04}. However, several lines associated to \fei\ multiplets 1 and 9 and opt 30 have been convincingly identified in a strong \feii\ emitter \citep{grahametal96}, and the emission of \ion{Fe}{1} with respect to \ion{Fe}{2}  might increase with metallicity \citep{sigutetal04}. The issue of \ion{Fe}{1} emission deserves further investigation not last because \ion{Fe}{1}  intensity is strongly dependent on the assumed \feii\ model at 2900 -- 3000 \AA. For the sake of present paper we checked that the peak shift of the \mgii\ is basically unaffected even by strong changes in the assumed \ion{Fe}{1} strength.}

\section{Results}
\label{results}

\subsection{Broad Line Profile Analysis}
\label{profan}

\begin{figure}
\includegraphics[scale=0.35]{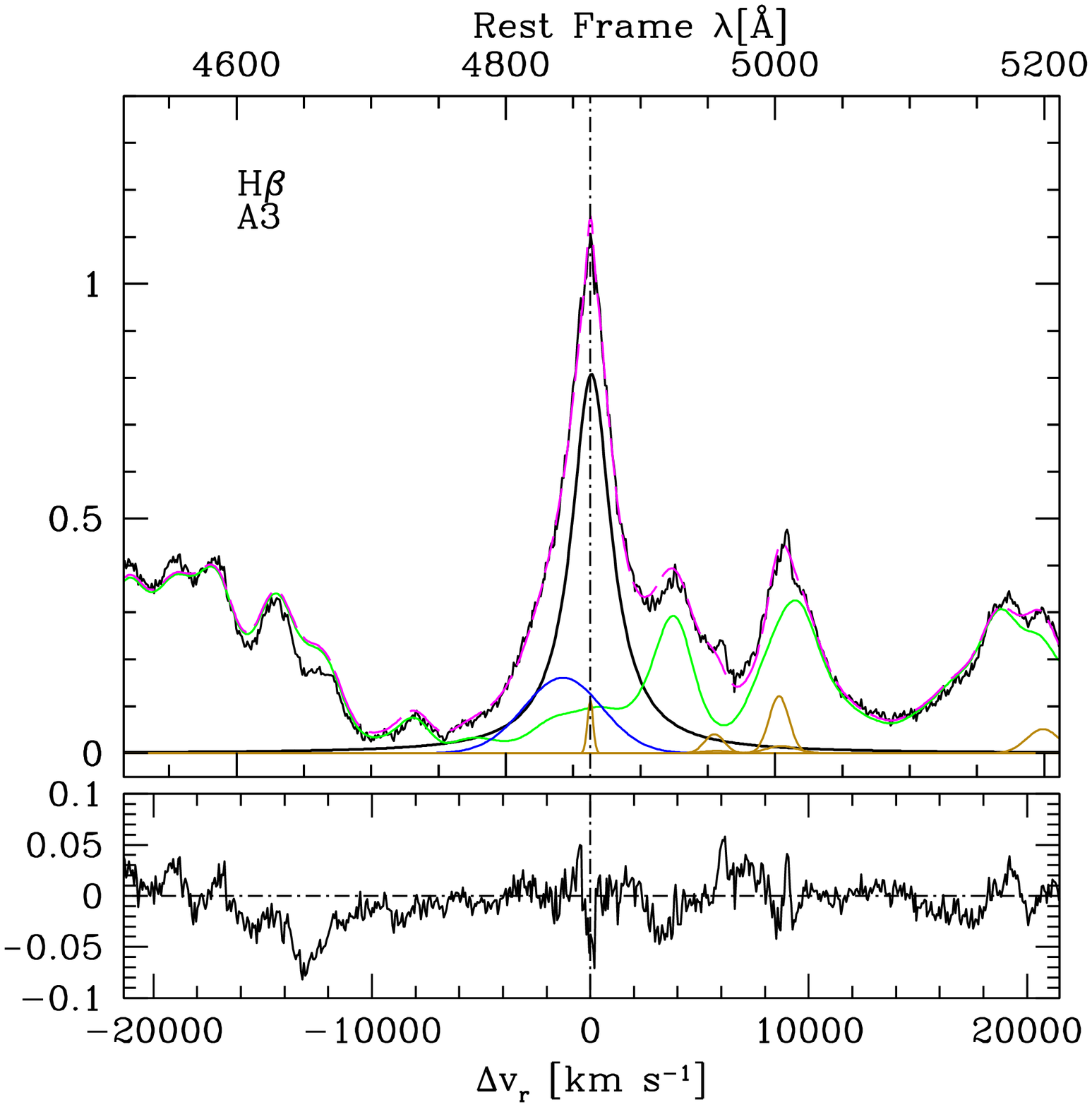}
\includegraphics[scale=0.35]{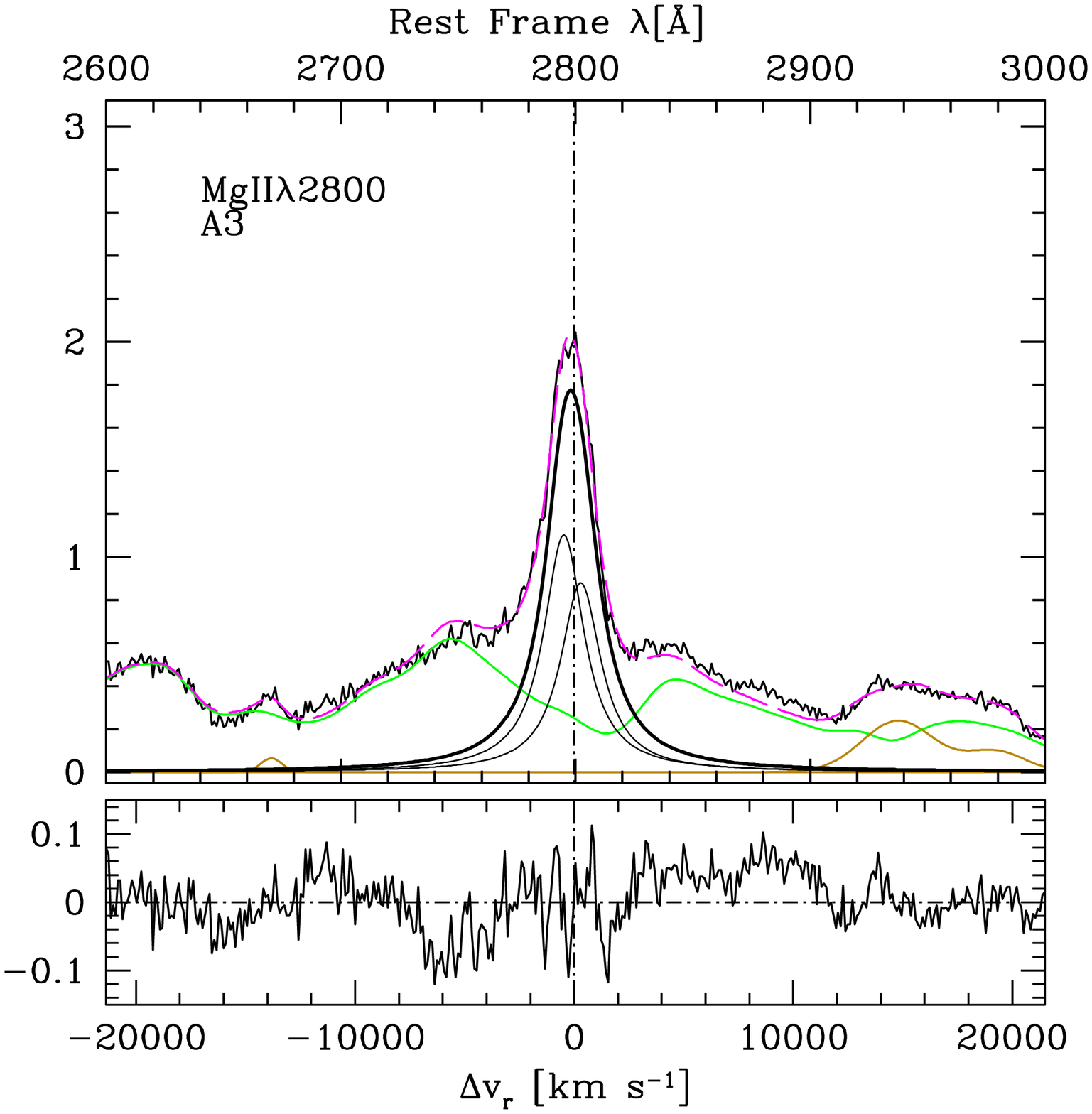}\\
\includegraphics[scale=0.35]{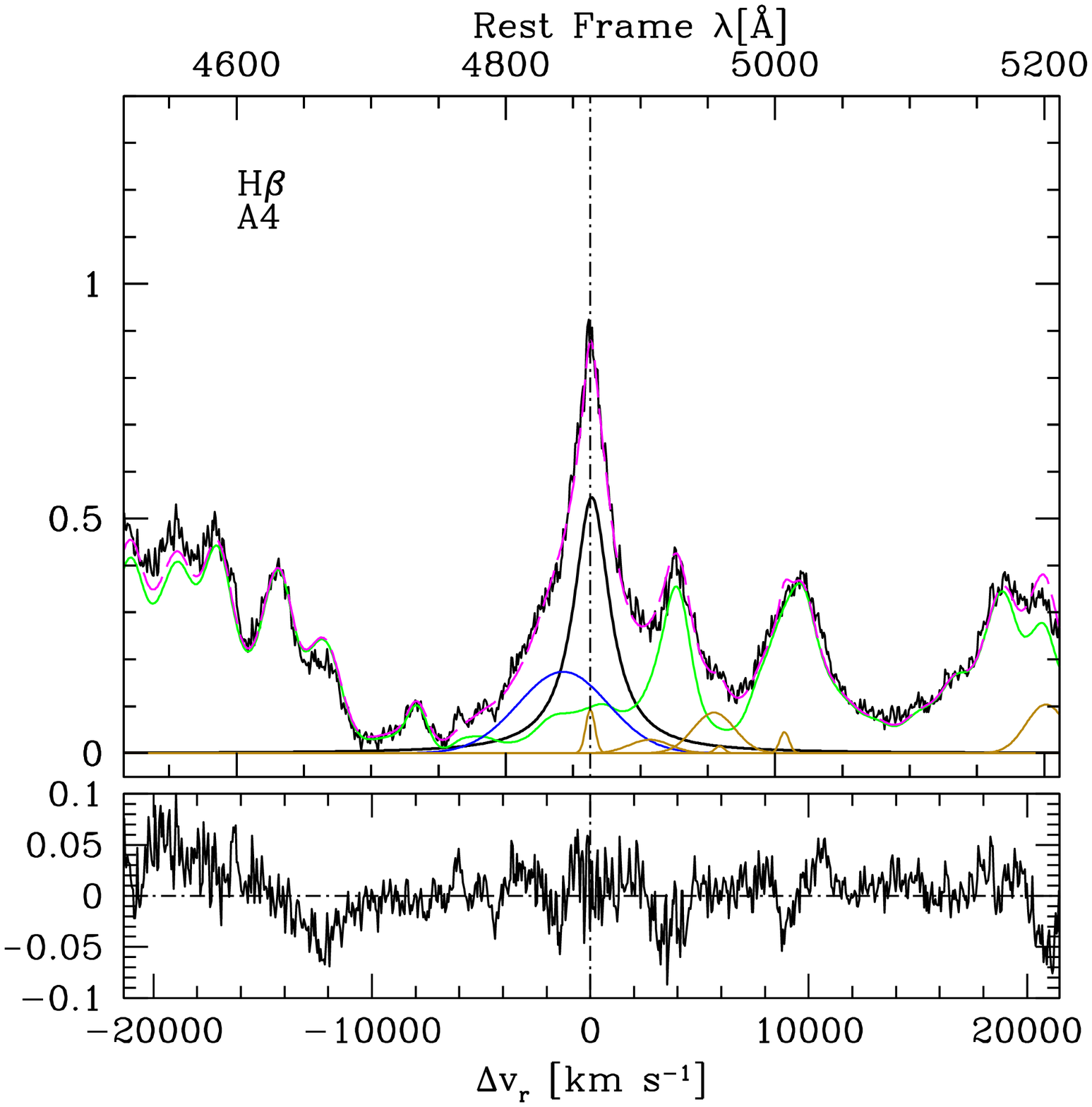}
\includegraphics[scale=0.35]{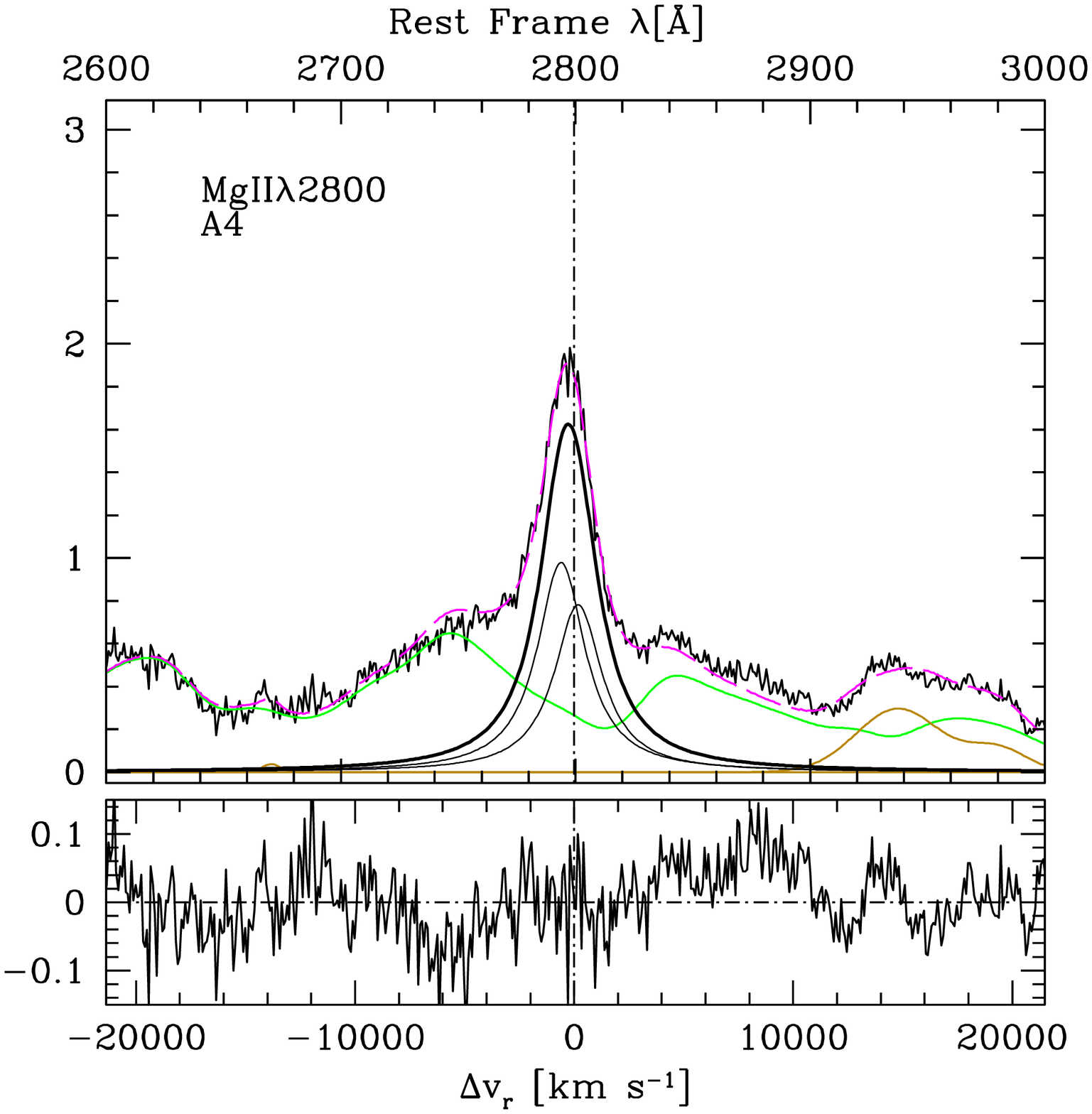}
\caption{Spectra of H$\beta$ (left panels) and \mgii\ (right
panels) for spectral types
A3 (top) and A4 (bottom). The right panels show  a model  that
assumes a shifted BC for \mgii\  and
an almost unshifted BC + a blueshifted Gaussian component  (thick blue line)  for \hb.  Meaning of all other symbols is the
same as Figure \ref{fig:hbmga1a2}.  \label{fig:hbmga3a4}}\end{figure}

\begin{figure}
\includegraphics[scale=0.35]{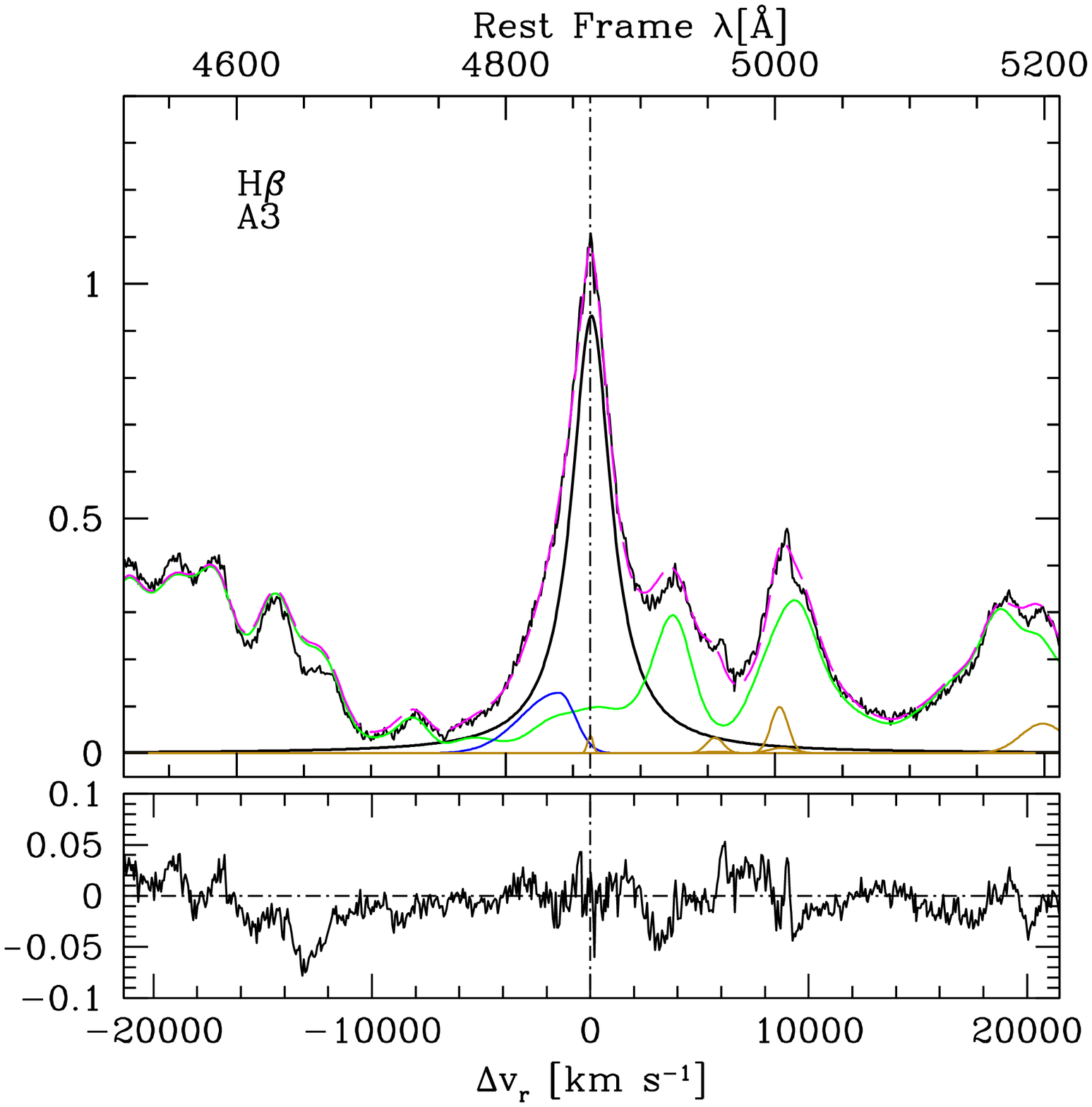}
\includegraphics[scale=0.35]{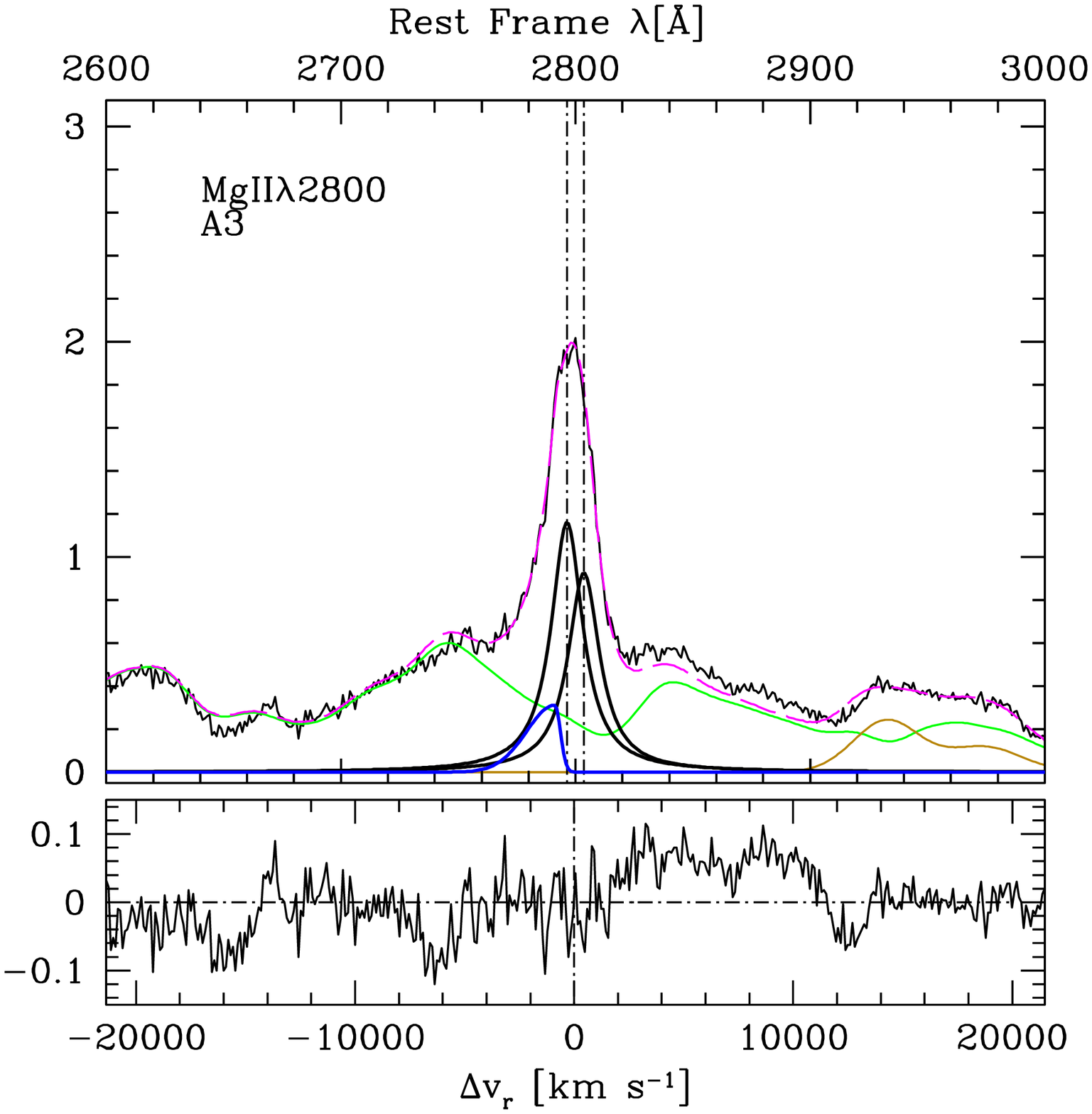}\\
\includegraphics[scale=0.35]{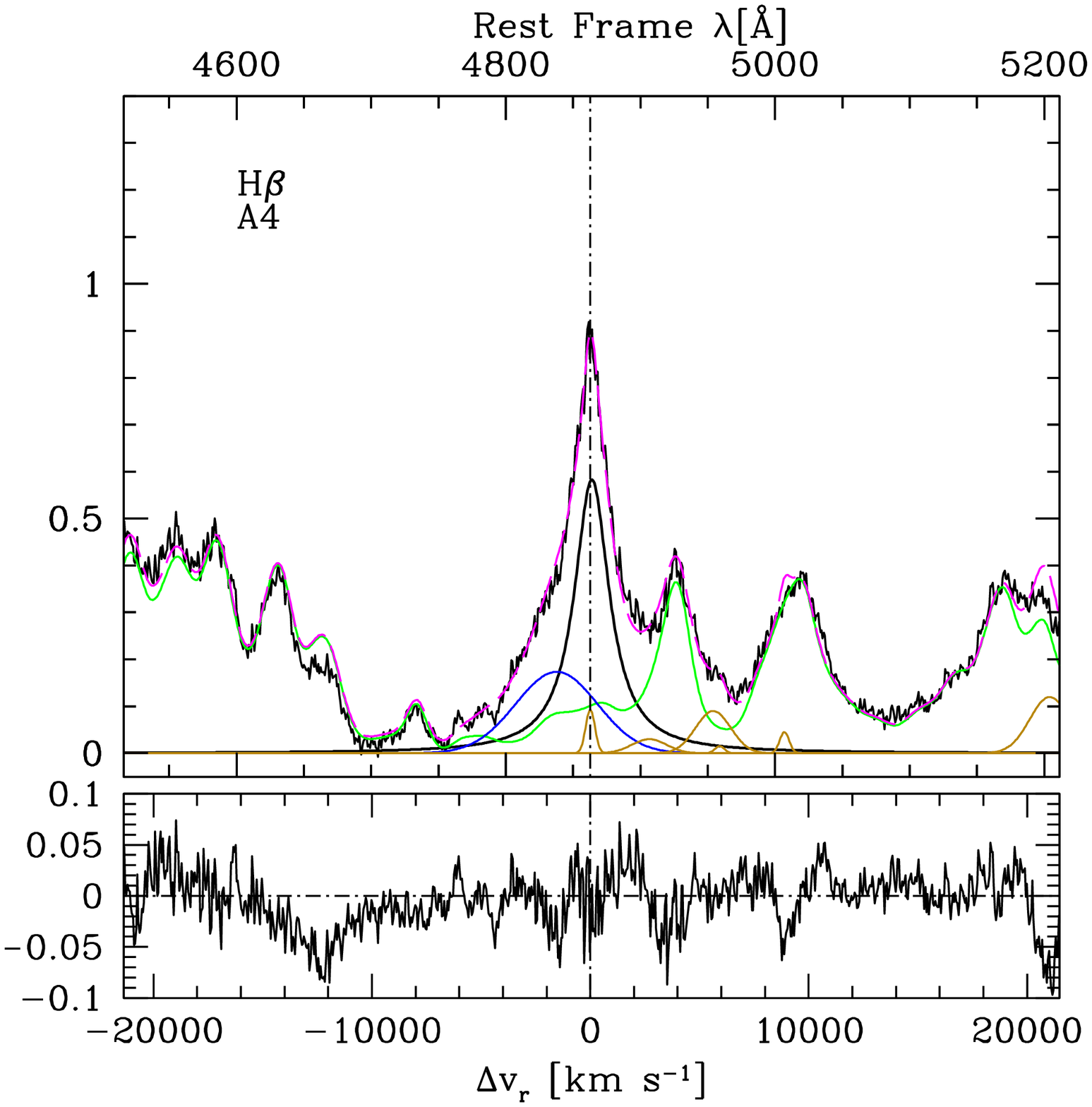}
\includegraphics[scale=0.35]{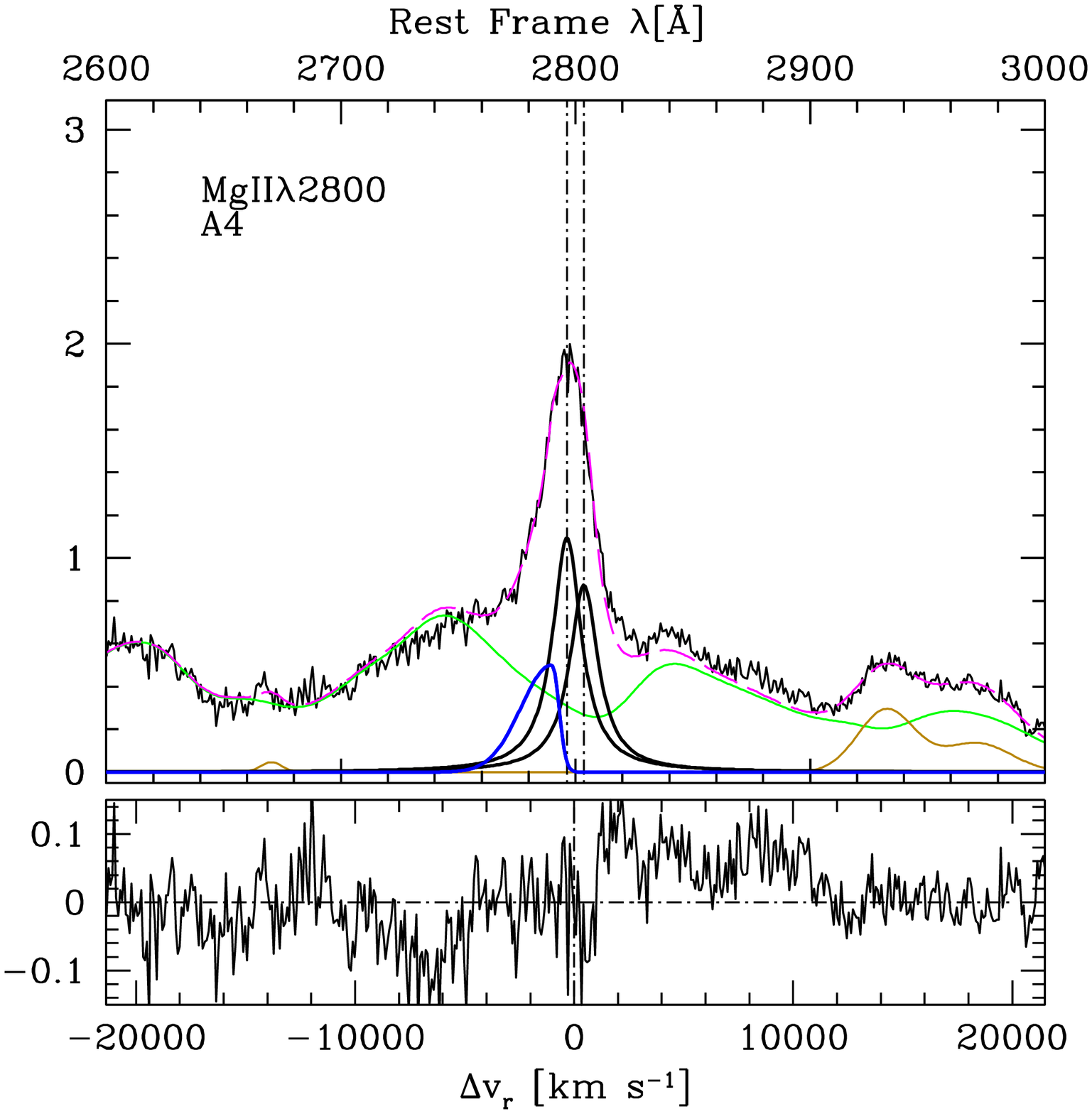}\\
\caption{{ Spectra of \hb\ (left panels) and \mgii\ (right
panels) for spectral types A3 (top) and A4 (bottom), as in the previous figure.  
The  panels show   models of both \hb\ (left) and  \mgii\  that
assume an (almost) unshifted BC and
a blueshifted skewed Gaussian   component.  Meaning of symbols is the
same as Fig. \ref{fig:hbmga3a4} except
for  the individual \mgii\ components that are shown by thick lines in the
right panels. Refence dot-dashed lines are drawn at the vacuum laboratory wavelength of the two \mgii\ components. }
\label{fig:hbmga3a4skew}}\end{figure}

Composite spectra were analyzed using {\sc specfit} with $\chi^{2}$ minimization techniques appropriate for  non-linear multi-component fits \citep{kriss94}.  The procedure allows for simultaneous continuum, \feii\ and  narrow line fitting. Two \feiiuv\ emission templates were applied: the theoretical one by \citet{bruhweilerverner08}, and an empirical template produced by \citet{tsuzukietal06}.
% provide \feiiuv\ emission templates computed from  {\sc cloudy} simulations using an 830 level model of the Fe+ ion. 
Use of two independent templates was justified by possible effects that \feiiuv\ subtraction might have on the measurements of line shifts. Continuum subtracted composite spectra are shown in  Fig. \ref{fig:hbmga1a2} for bins A1 and A2, and  in Fig. \ref{fig:hbmga3a4} and Fig. \ref{fig:hbmga3a4skew} for bins A3 and A4. The \mgii\ doublet was first modeled as two Lorentzian-like functions of same width and  relative intensity ratio 1.25:1. { This assumption is justified by the value of the prototypical A3 source I Zw1, and by the physical conditions within the BLR \citep{laoretal97a}. { It is possible that the lines become fully thermalized at the extreme optical depth of the low-ionization line (LIL) BLR, justifying the assumption of a 1:1 ratio}. We have carried out several  fits
also for the 1:1 case but the results on line shifts and widths are very close to the 1.25:1 case and do not affect any of the conclusions discussed below.}  

{ Table \ref{tab:specfit} reports the results of the our multi-component analysis:  intensity, shift $\pm$\ uncertainty at 2$\sigma$ confidence level\ (estimated as described at the end of \S \ref{sample}), and FWHM of the two line components used to model line profiles: the broad component (BC) and, when appropriate, a blueshifted component (blue). % {Peak \mgii\  \vr\ $2\sigma$ uncertainty is $\approx 50$\ \kms\ for sp. types A1,A2,A3 and $\approx 85$\ \kms\ for A4. } 
}  {  The radial velocity was measured with reference to the vacuum wavelength of the $^{2}P_{\frac{3}{2}} \rightarrow ^{2}S_{\frac{1}{2}}$\ component, 2796.35 \AA.} Uncertainties { in BC centroid shifts} are significantly larger toward the line profile base than at peak, and are estimated to be $\pm$ 400 \kms\ for the centroid at 1/4 peak intensity. {  Columns BC and blue Int.  report  \hb\ and \mgii\  line intensity  normalized by   continuum flux at 5050 \AA. The normalization at 5050 \AA\  ensures that  the \hb\  value roughly corresponds to the line equivalent width,  and that meaningful intensity ratios \mgii/\hb\ can be computed from the values reported in the Table. Note that \mgii\ intensity values are for the doublet, while reported FWHMs are for an individual component.  The formal uncertainty in FWHM measurements of \hb\ and \mgii\ BC is, in absence of systematic effects, around 2\%\  for spectral types A1 and A2, and $\approx$ 5\% in all other cases. Shifts and width of the blue components are subject to considerable larger uncertainties since they are close to the much stronger BC. In addition their values depend on the line profile assumed for the fit. Formal uncertainty (i.e., without considering the possibility of different profile shapes) derived on the FWHM of the blue components is $\ltsim$ 10\%\ in  all cases.  }

Fits to H$\beta$\ and \mgii\ in A1 and A2 bins needed only  symmetric, unshifted  Lorentzian line components.  The BC accounts for the entire \hb\ and \mgii\ profiles in A1 and A2 sources where the ratio FWHM(\mgii) / FWHM(\hb)$\approx$ 0.75 -- 0.80 (Tab. \ref{tab:specfit}). This FWHM ratio also holds for  Pop. B sources and, therefore, for 90\%\ of quasars   \citep[][Sulentic et al. 2012, in preparation]{wangetal09,trakhtenbrotnetzer12}. {  The simplest interpretation is that the emissivity weighted distance of the \mgii\ emitting gas is somewhat larger than that for \hb ($\approx$ 1.5 following
the virial assumption, see also \S \ref{discussion}).  

The  \hb\ profiles in A3 and A4 bins also involve an (almost) unshifted, symmetric component (the broad component  BC) with FWHM(\hbbc) $\sim$ 2000 \kms\ which we assume to be the virial broadening estimator. However, fits to the H$\beta$\ profile for sources in A3
and A4 bins  require an additional blueshifted (Column ``blue'' in Tab.
\ref{tab:specfit}) component in order to minimize residuals 
\citep[e.g.][]{leighly00,leighlymoore04,marzianietal10,wangetal11}.  The
``blue'' component has been modeled  first as a symmetric Gaussian, in line with past work.
The right panels of Fig. \ref{fig:hbmga3a4} show that a good fit to the \mgii\ profiles in A3 and A4 is possible using a shifted symmetric Lorentz function with a profile shift of a few hundred \kms.  

At a second stance    a skewed Gaussian \citep{azzalini85} has been considered
 for both \hb\ and \mgii\ (Fig. \ref{fig:hbmga3a4skew}).  In this case, a two-component model is possible also for the spiky \mgii\ profile.  We assumed unshifted BC \mgii\ emission with FWHM(\mgii) = 0.8 FWHM(\hb),  plus an additional  blueshifted component described by a skewed Gaussian as for \hb. The resulting \mgii\ line decomposition is shown in the right panels of  Fig. \ref{fig:hbmga3a4skew}. Line parameters are reported in Tab. \ref{tab:specfit} (A3b and A4b). }    

\subsection{A systematic \mgii\ Blueshift}
\label{blueshift}

Bins A3 and A4 (10\%\ of all quasars) behave differently than the wide majority of quasars since their FWHM(\mgii) $\geq$ \hb. The ratio FWHM(\mgii)/FWHM(\hb)  is larger than unity in bin A4, with \hb\ and \mgii\ showing the same width in bin A3 (lower panel of Fig. \ref{fig:trenda}) where a peak blueshift is already highly significant. As mentioned, the  \mgii\ doublet appears blueshifted with respect to rest frame and \hb\ in spectral types A3 to A4 where \hb\  shows evidence for a blueshifted component (Fig. \ref{fig:hbmga3a4}; see also  \citealt{marzianietal10}).  The \mgii\ blueshift reaches $\approx$ 20\%\ of the half line width in bin A4. In bin A3 and A4 the core of the \mgii\ profile is narrow enough to appear visually displaced relative to the rest frame (Fig. \ref{fig:hbmga3a4}). { Even measuring the position of the broad-line core (without any correction because of contaminant lines) with {\sc splot} we obtain consistent values. } The effect is too large to be ascribed to sources of uncertainty on rest frame, and it is even more significant if relative line shifts are considered.   { As mentioned, we repeated the fits for all bins assuming that  the doublet ratio is 1.0:1.0. This  results in a slightly larger \mgii\ blueshift.  To further test the reality of the shift, we considered that  the maximum doublet ratio  for expected physical conditions  in the BLR is 1.5:1.0. We   constructed noiseless mock profiles to derive a peak wavelength in case \mgii\ treated as single line, for doublet ratios 1.5,1.25,1 to 1. The effective wavelength of the doublet is 2799.1, 2799.4, 2800.1 for the three ratios respectively. Also in the case 1.5:1.0, the peak shift will remain significant. 

The \mgii\ A3 and A4 fits with a shifted symmetric function  probably yield  only a lower limit to the shift amplitude since, if a two component interpretation is correct, they include also unshifted emission line gas.  Yet, these \mgii\ fits are meaningful since they provide a robust measurement of a significant blueshift affecting the \mgii\ line profile.  } {  Renouncing the symmetric Gaussian approximation for the blueshifted emission (in \hb) provides support for profile decomposition into two components (Fig. \ref{fig:hbmga3a4skew}). With the  exception of  A4 \hb,  both \hb\ and \mgii\ median spectra show  blue components with  strong blueward asymmetry: the blue component profiles  vaguely resemble the ``trapezoidal'' shape of the    I Zw 1  \civ\ profile \citep[c.f.][]{leighly00}.  The lower \vr\ derived for \mgii\ blueshifted component with respect to the one of \hb\ in bins A3 and A4 is consistent with the profile shape difference since \hb\ is more affected toward the line base, while \mgii\ is affected to closer to the line core.   Indeed, in all cases the shift of \hb\ is larger than the shift of \mgii. This result provides an important constrain on the emitting region structure (\S \ref{discussion}). Considering the strongly skewed blue component line profiles in A3 and A4, the estimated shifts reported in Tab.\ref{tab:specfit}  could be considered more properly as  upper limits since symmetric blueshifted Gaussians would yield some emission under the BC with lower shift values. }

\begin{figure}
\includegraphics[scale=0.4]{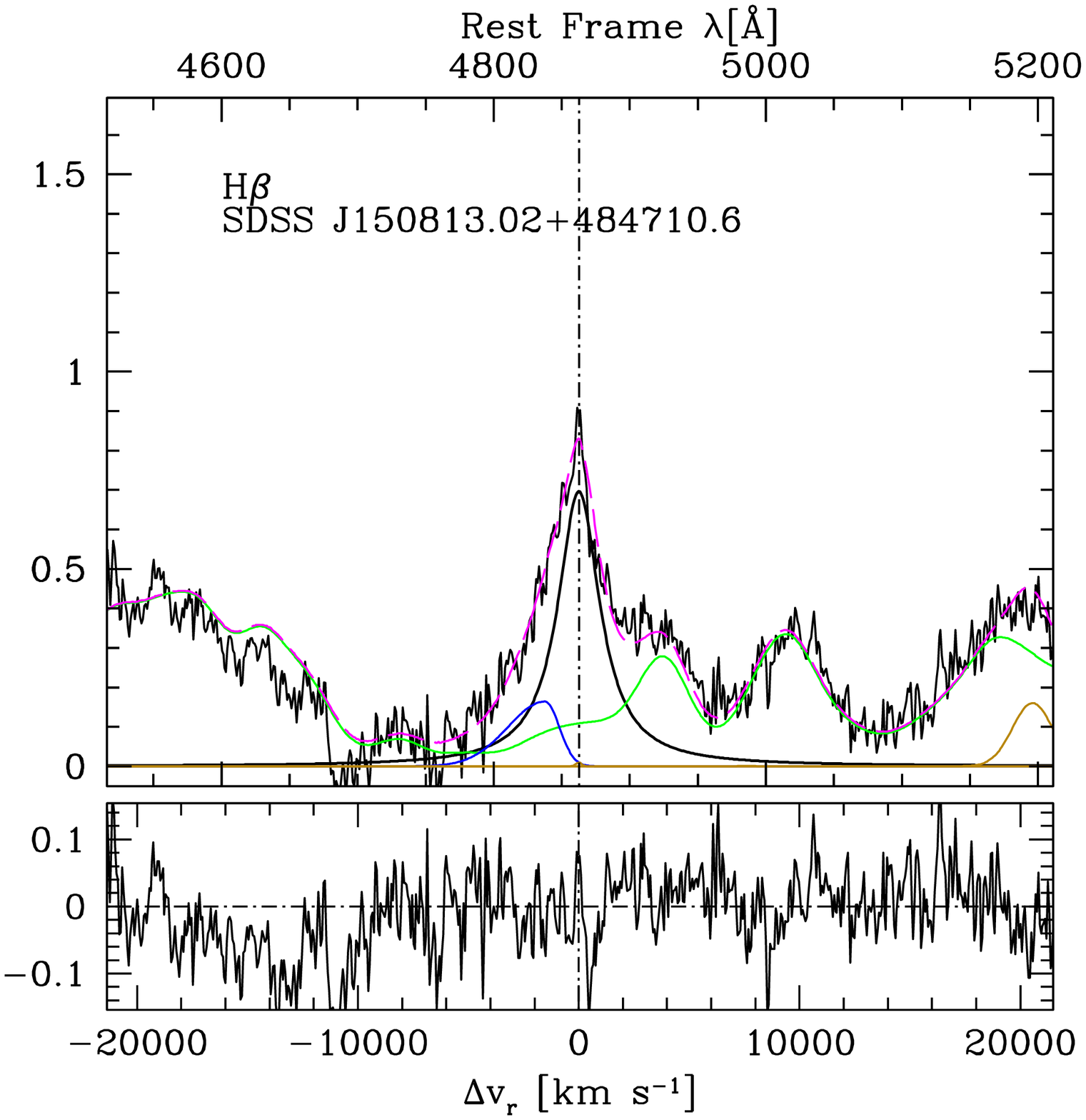}
\includegraphics[scale=0.4]{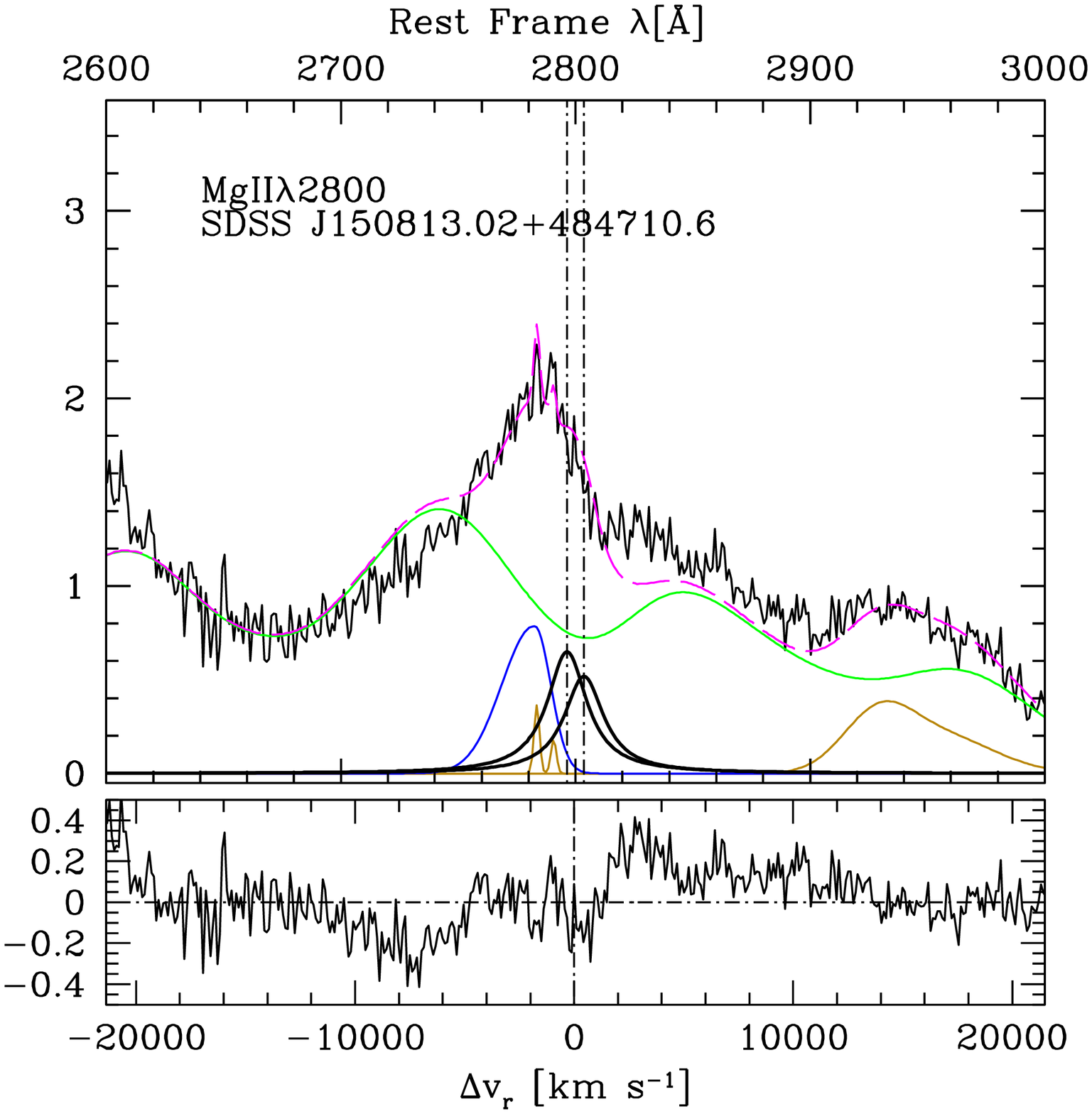}
\caption{Spectra of H$\beta$ (left panels) and \mgii\ (right panels) for
quasar SDSS J150813.02+484710.6. Vertical scale is specific flux in units
of 10$^{-15}$  \ergss\ \cmq\ \AA$^{-1}$.  The meaning of all other symbols is
the same as  Figs. \ref{fig:hbmga1a2} and \ref{fig:hbmga3a4skew}. 
\label{fig:sdss}}
\end{figure}

The most extreme source  in our sample involves SDSS J150813.02+484710.6 whose \mgii\ and \hb\ profiles are shown in Fig. \ref{fig:sdss}. \hb\ shows a prominent blue asymmetry while \mgii\ is fully blueshifted with width and shift amplitude similar to the ones measured for \civ\ in extreme Pop. A sources such as I Zw 1. For this source we apply to the \mgii\ profile only the profile model appropriate for \civ\ of extreme Pop. A quasars i.e.,   an unshifted double Lorentzian + a blueshifted component approximated with a skewed Gaussian.   { While uncommon, other sources like J1508+48 have been found  \citep[see e.g., Q1258+1404; ][]{bartheletal90}.}

\subsection{Major trends}

\subsubsection{Spectral types}

\begin{figure}
\includegraphics[scale=0.85]{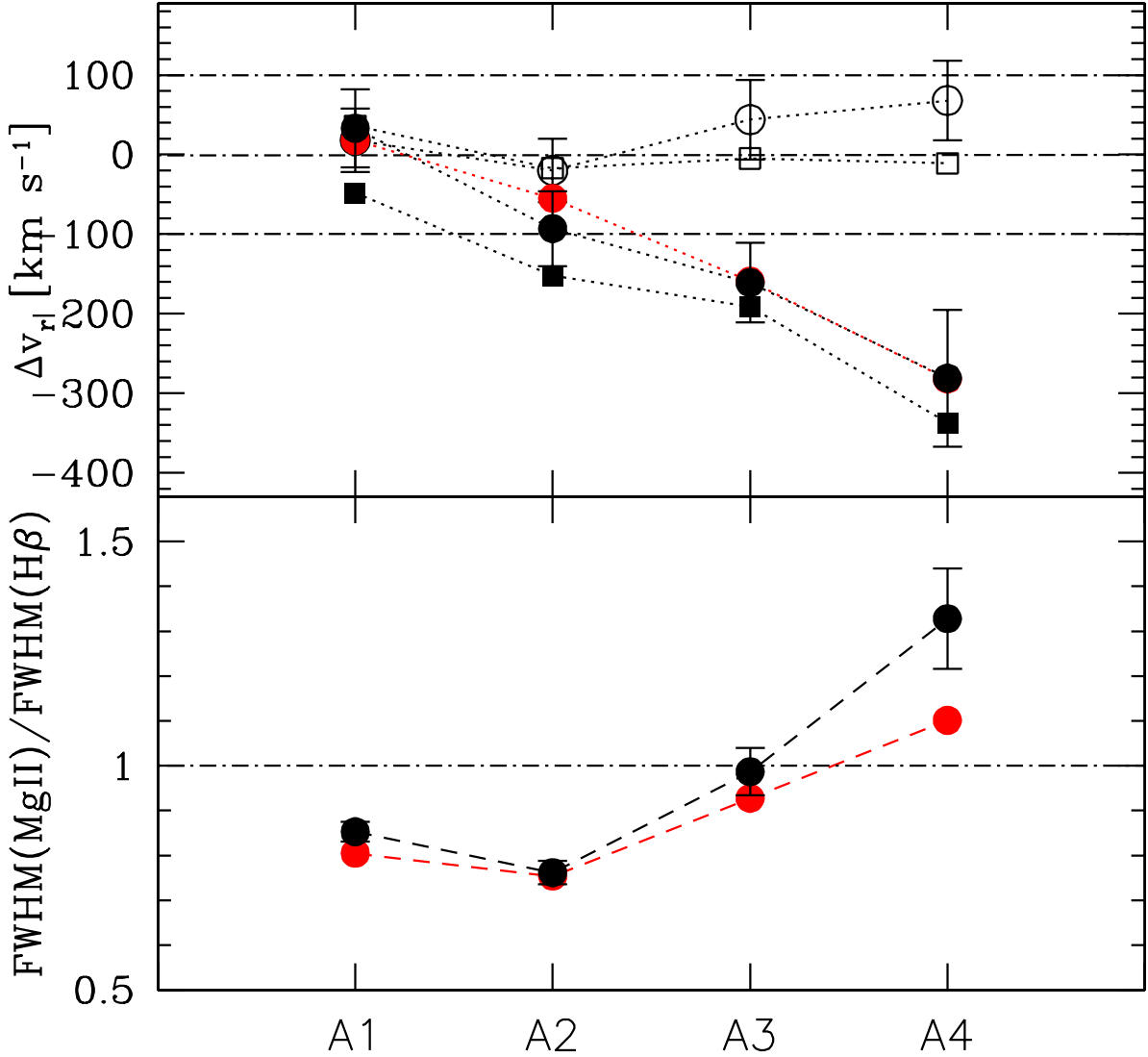}
\caption{Trends in Pop. A as a function of spectral type. Filled circles:
\mgii; open circles: \hb. In the upper panel
the BC shifts are relative to rest frame. Error bars are at 2$\sigma$\ confidence level, and are shown only for one measure of \hb\ and one of \mgii\ for clarity. The squares indicate an additional measurement of the core
centroid of the line obtained after
removal of NC emission.   The red circles refer to measurements obtained
applying the \feii\ template of \citet{tsuzukietal06} in the region of
\mgii. The dot dashed lines
at $\pm$100 \kms\ in the top panel
defines a zone where  peak measurements are consistent with zero shift (at
2$\sigma$ confidence level).  In the lower panel the ratio
FWHM(\mgii) (individual component) over FWHM(\hbbc)
is plotted.   Error bars are again at 2$\sigma$\ confidence level.  \label{fig:trenda} }
\end{figure}

The upper panel of Fig. \ref{fig:trenda} shows radial velocity \vr\ trends
of broad component peaks as a function of spectral type, while the lower panel shows trends of the
ratio FWHM(\mgii) / FWHM(\hb). \mgii\ line shift measurements shown in Fig. \ref{fig:trenda} were carried out in three different ways:
(1) using {\sc specfit} with the theoretical \feiiuv\ template, (2) using {\sc specfit} with the
\feiiuv\ template from \citet{tsuzukietal06}
and (3) measuring the position of the broad-line core without any
correction. The three sets of measures
yield consistent trends. The interesting change in the lower panel of the
figure involves a tendency for the
FWHM ratio to increase first to parity in bin A3 and finally to
FWHM(\mgii) $>$ FWHM(\hb) in bin A4 \citep[cf.][]{trakhtenbrotnetzer12}.
This trend likely accounts for the large scatter (and convergence toward
parity) of single-source  measures in Figure 2 of \citealt{wangetal09}). The
consistent behavior of both shift and FWHM strengthens our confidence that the trends are real. The increase in FWHM(\mgii) lends support to the hypothesis that an additional blueshifted component is emerging  on the blue side of an unshifted BC with
FWHM(\mgii)$\approx$ 0.8 FWHM(\hb). { A symmetric blueshifted Lorentzian
model is helpful for ascertaining the reality of the blueshifts  but seems physically unrealistic. The
blueshifts are more likely associated with a \mgii\ component due to gas moving at larger velocity than the ones inferred from the global shift of the line core (Tab. \ref{tab:specfit}).}

%\footnote{The FWHM ratio is the most unbiased estimator, since it is independent from the actual distribution of FWHM(\hbbc). 
%There is however no known  bias affecting the original survey and leading to a systematic loss of  broader sources in bin A3 and A4.}  
%\footnote{In the following we report values for spectral type A4, but FWHM(\hbbc), $L$,  and $L$(\mgii) are similar for A3 and A4 as well as for SDSS J1508+4847 and lead to analogous inferences unless otherwise noted.} Fig. \ref{fig:trendalla} shows the behavior of line shifts as a function of Eddington ratio. 

\subsubsection{Eddington ratio}

\begin{figure}
\includegraphics[scale=0.85]{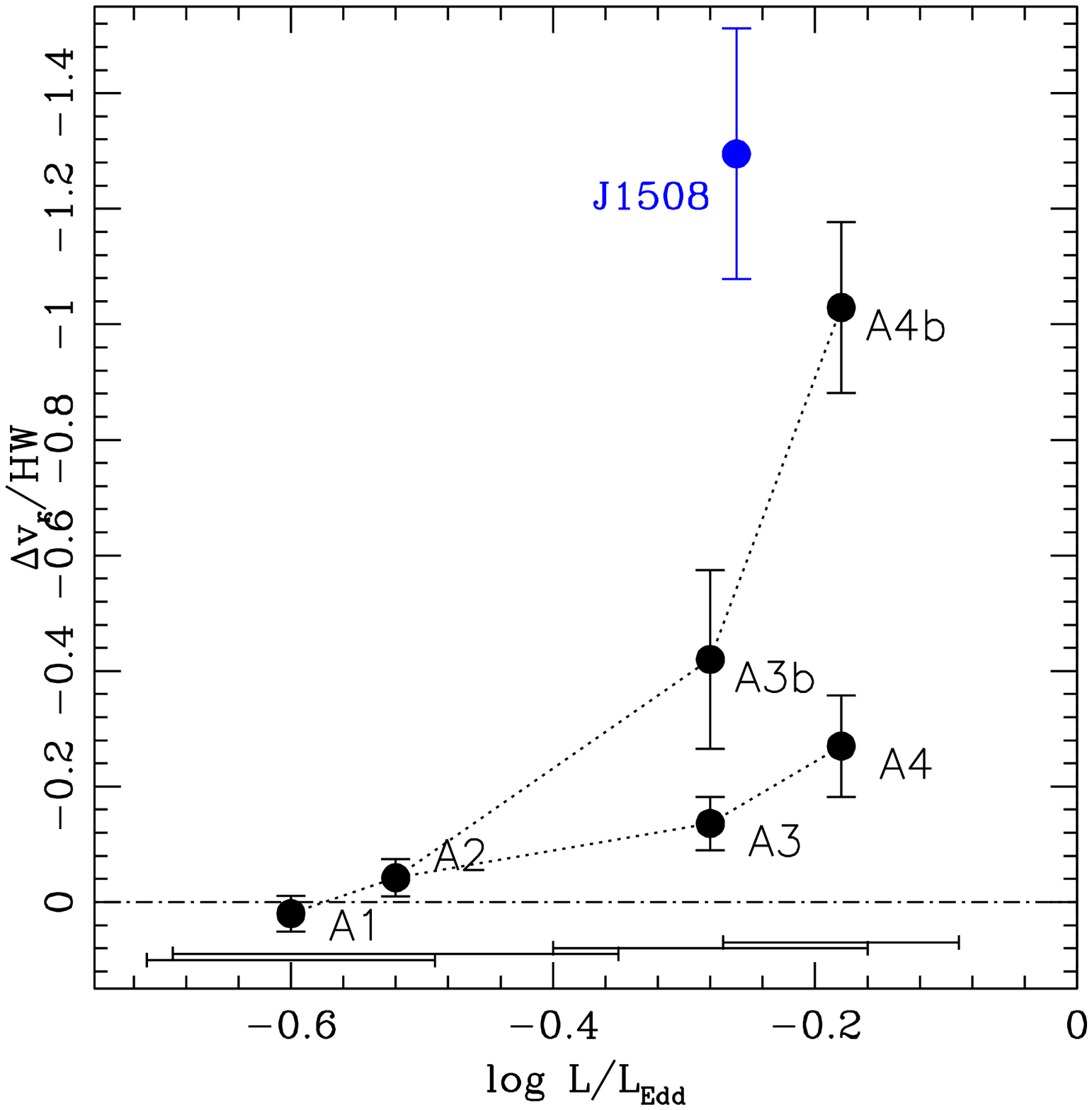}
\caption{{  The ratio between the peak \mgii\ shift and \hb\ half-width
half maximum (HWHM) as a function of
Eddington ratio. Data points are shown for an unresolved \mgii\  fit (A1,
A2, A3, A4), as well as for the
blueshifted component in two component fits (A3b and A4b).  Error bars are
at 2$\sigma$\ confidence level.  \label{fig:trendalla} SIQRs in \lledd\
values are $\approx 0.1$ for bins A1,A3,A4, and  $\approx 0.16$ in bin A2.
Eddington ratio ranges  corresponding to $\pm$SIQR are indicated in the
lower part of the diagram for all four spectral types. The blue data point
refers to
a two component fit for source SDSS J150813.02+484710.6. }}
\end{figure}

We estimated  median \mbh\ values   {  computing median 5100 \AA\ luminosities from fluxes of all sources in each bin }and with FWHM measures of the median composites following  the  prescription of \citet{assefetal11}.   A bolometric correction
to the 5100 \AA\ luminosities was applied following \citet{nemmenbrotherton10} in order to derive \lledd\ values.
{ The semi inter-quartile range  (SIQR) of \lledd\ has been estimated using
the  individual FWHM \hb\ measurements carried out for spectral bin assignment (\S \ref{sample}).}
All bins show very similar median bolometric luminosities $\log L\approx 46.2$\ [\ergss].  Fig.
\ref{fig:trendalla} shows the peak shift  of
the \mgii\ broad and blue components as a function of \lledd. Shift values
have been  normalized  by the \hb\ half-width at half-maximum (HW) in
order to provide an indicator  of dynamical significance  to the shifts. 
The peak \mgii\ shift
is consistent with zero for spectral types A1 and A2 but begins to appear
at $\log$\ \lledd$\sim$ --0.5 at  type A2 increasing to 0.2/0.3 of
half width for A3 and A4. {  The shift amplitude is much larger when the blueshifted component is considered. 
 If 0.8 $\cdot$ HW can
be considered as a rough estimator of the virial velocity of the \mgii\
emitting gas then A4 and SDSS J1508+4847 show outflows close to escape
velocity. The main difference between median A4 and the source  SDSS
J1508+4847 is related to the amount of outflowing gas: in A4 the blue
component is contributing 1/7 of the total line emission while it exceeds
1/3 in SDSS J1508+4847.}  

\section{Discussion}
\label{discussion}

%Results from this and previous low redshift samples  suggest that Pop. A sources show systematically  higher \lledd\ than pop. B sources, with A3 and A4  as the loci of most extreme  \lledd. These results assume that \hb\ is the most reliable virial estimator. Even in bins A3 and A4 there is a strong symmetric \hb\ broad  component  that serves as a virial broadening estimator. \mgii\ shows a systematic blueshift that challenges its  status as a valid virial estimator. This shift involves the entire \mgii\ core unlike the situation for \hb.  Sources in bin A3,  and A4 are the ones showing the largest \mgii\ blueshift. SDSS J150813.02+484710.6 shows the largest blueshift ever found, with a broad and blueshifted profile that resembles the one of \civ\ in some NLSy1s. 
%\footnote\{Since $\log L_\mathrm{bol} \approx {46}$ \ergss\ with a sample variance of $\pm 0.2$ in each bin, maximum \lledd\ corresponds to minimum \mbh} 

The \mgii\ profile in median spectra is different from both \civ\ and \hb.
Comparison between the latter two lines yielded
inferences about the BLR  structure \citep{marzianietal96}. The \civ\ blueshift in low $z$
sources is thought to be associated with a wind
component whose prominence increases with \lledd\ along the 4D eigenvector
1 (4DE1) sequence. The \civ\ profile in bins A2-A4 can be modeled as a combination of the \hb\ BC profile plus a fully blueshifted component that  accounts for most of the flux. This basic 
scenario has been confirmed by recent work (see e.g. \citealt{richardsetal11,wangetal11}). In the case of \mgii\ we see a \vr\ displacement that is also significant with  200 -- 300 \kms\ if the shift is measured on the full profile. As pointed out, this is likely a lower limit. If the blue component is considered,  the shift amplitude is much larger ($\sim$ --1000 \kms) but  significantly lower in \mgii\ than in \hb.  We are considering median spectra so the profiles represent the median behavior of line profiles -- in some sense equivalent to a single source rms profile.  So we can ask: what is the typical relation between the  blueshifted \civ\ and \hb\ on the one hand  and the blueshifted \mgii\ emission on the other? 

\subsection{\mgii\ blueshift: emission from a radiation-driven wind}

The \mgii\ blueshift is most straightforwardly interpreted as due to outflow motions of the line emitting gas with preferential obscuration of the receding part of the flow. The fact that large blue shifts are observed when \lledd\ is highest indicates a role of radiation force in acceleration of the gas. If we consider  gravitation and ionizing radiation as the only forces (neglecting drag forces and pressure gradients)  the radial acceleration can be written as $a(r) \propto \sigma M_\mathrm{BH} /r^{2} \left[\alpha(r)/(\sigma N_\mathrm{c}) L/L_\mathrm{Edd} - k\right]$, where $\alpha(r)$ is the fraction of bolometric luminosity absorbed, \nc\ the column density and $\sigma$ the Thompson scattering cross-section, { and $k$\ a constant term}. If the first term in square brackets exceeds $k$, the outflowing velocity field of the gas will follow the form $v(r) = v_\mathrm{t}\sqrt{1- r_\mathrm{min}/r}$, where $r_\mathrm{min}$\ is the launching radius of the wind, and $v_\mathrm{t}$\ the terminal velocity ($\propto \sqrt{L/r_\mathrm{min}}$).
Outflows driven by line and/or ionizing photon pressure  can accelerate the line-emitting gas to $v_\mathrm{t} \approx k' \left({\cal M} { L}/{r}\right)^{\frac{1}{2}} \approx \left({\cal M} { L}/{L_\mathrm{Edd}}\right)^{\frac{1}{2}} v_\mathrm{Kepl}$\ where $v_\mathrm{Kepl}$\ is the Keplerian velocity at the launching radius of the wind 
%and $k = (\sigma_\mathrm{T}/4 \pi m_\mathrm{P})^{1/2}$\
 \citep[e.g.][]{laorbrandt02}. In the case of outflows driven by ionizing radiation the force multiplier $\cal{M}$\ is expected to be: ${\cal{M}} = \frac{\alpha}{\sigma N_\mathrm{c}} \approx 7.5 \alpha_{0.5} N_\mathrm{c,23}^{-1}$  where $\alpha \approx$0.5 for Compton-thin gas optically thick to the ionizing continuum \citep{netzermarziani10}. ${\cal{M}}$ can be $\gg 1$ in the case of line driven winds \citep{progaetal00}. 

The largest \civ\ blue shifts are observed in spectral types A3 and A4 (in A3, $ -1000 \ltsim$   $\Delta v_\mathrm{r} \ltsim -2000$ \kms) where also the  \mgii\ shifts occur. It therefore seems unlikely that there is no connection between \civ\ and \mgii: both lines may be emitted as part of the same flow. However, bulk emission are expected to occur at different distances and/or in different physical condition. In the framework of photoionization, \mgii\ emission is associated with low-ionization  and relatively large column density gas \citep{netzer80,koristaetal97}. Within a gas slab or cloud, \mgii\ is emitted mainly beyond the fully ionized zone of geometrical depth $h \sim 10^{23}U$\nh$^{-1}$\ (with $U$\ being the ionization parameter)  where all of the  \civ\ is emitted.  The total column density needed for substantial \mgii\ production is not well constrained at very low ionization, since the fully ionized zone is already a tiny fraction of the emitting gas slab  if \nc $\sim 10^{23}$ \cmq\ and $\log U \ltsim -1$. 

Since $v_\mathrm{t} $\ is  proportional to both \nc$^{-1/2}$  and $r_\mathrm{min}^{-1/2}$,  \mgii\ emission may occur at higher column density and/or larger distance than blueshifted \civ. 
A large $r_\mathrm{min}$\ is consistent with the overall
symmetry of the \mgii\ profile base (\S \ref{blueshift}; Figs. \ref{fig:hbmga1a2}, 
\ref{fig:hbmga3a4}, \ref{fig:hbmga3a4skew}).  Reverberation mapping indicates
that high ionization lines are emitted closer to the central continuum
than low-ionization lines \citep[e.g.][]{petersonwandel99}. It is possible
to ascribe the blueshifted \mgii\ emission entirely to a larger radial
distance  if \mgii\ arises  $\sim {2}$ times more distant than \civ. On the other hand, if
\lledd\ $\rightarrow$ 1,  $v_\mathrm{t} \approx \sqrt{\cal M}
v_\mathrm{vir}$. Restricting our considerations to order-of-magnitude estimates due to
uncertain shift values, a reasonable increase in column density over the
standard value   \nc $\sim 10^{23}$ \cmq\ would also suffice to reduce the
\mgii\ shifts close to the observed values.

An assumption from the behavior of the FWHM ratio is that the \mgii\ flows start close to where the bulk of the low-ionization lines (\hb\ in the present case) is emitted. { The following considerations
apply also if FWHM(\mgii)/FWHM(\hb)$\approx$ 0.8 implies
$\approx$1.5$\times$ larger emissivity-weighted distance  for \mgii\ than for \hb. }   The distance from the continuum source of the emitting gas $r_\mathrm{em}$ can be derived from the continuum luminosity at 5100 \AA\ following \citet{bentzetal09}: $\log r_\mathrm{em}\approx 17.6$ [cm] for spectral type A4. If $\log $\nh = 12 [\cm3], the ionization parameter is $\log U \approx -2.8$. The  \mgii\ luminosity is then $L($\mgii$) =  4\pi r_\mathrm{em}^{2} f_\mathrm{c} \sigma_\mathrm{em} = 4\pi r_\mathrm{em}^{2} f_\mathrm{c} \bar{\epsilon} \frac{N_\mathrm{c}}{n_\mathrm{H}}$, where $ f_\mathrm{c}$\ is the covering factor, $\sigma_\mathrm{em}$\  the emerging line flux for unit surface, and $\bar{\epsilon}$\  the  depth-averaged volume emissivity.   {\sc cloudy} \citep{ferlandetal98} simulations  indicate that $\sigma_\mathrm{em}$\       has  a minimum value $\approx 10^{7.8}$\  \ergss\ \cmq\ at $\log n_\mathrm{H}$ = 12
if $-2.8 \ltsim \log U \ltsim -2$, 10$^{23}$ \cmq $\ltsim$ \nc$\ltsim$10$^{25}$\cmq, $12 \le \log$\nh$\le 13$. In these ranges $ \sigma_\mathrm{em}$\ depends  slightly on \nc\ and $U$.
The computed $L$(\mgii) extrapolated to full continuum coverage is always
larger than the observed $L$(\mgii)  ($\approx 10^{43.5}$\ergss\ for A4).
This is the case if the geometry is assumed static and open or if a
velocity field appropriate for a wind (i.e., with photon local escape
probability following Sobolev's approximation) is considered. The derived
$ f_\mathrm{c} \ltsim 0.2$\   indicates partial covering of the continuum
as in a wind or in an ensemble of outflowing clouds. 

%A value \nc $\sim 10^{24}$\ \cmq\ is plausible in this simplified scenario.

Resonant line acceleration is expected to contribute to the dynamics of the flow in the physical scenario outlined above. Circumstantial evidence in favor of line acceleration is provided by the difference between a resonance UV line (\mgii) and a non-resonance line (\hb).  From the purely observational point of view, a line driven outflow would be convincingly demonstrated if the  ionizing photon flux were found unable to drive the line emitting gas to the observed outflow velocity or, in the context of absorption lines,  from ``line locking'' \citep[e.g.][]{cottisetal10}.  However, the ionizing photon flux appears to be  sufficient to accelerate the gas to the observed \mgii\ velocities and {   to escape velocity in  A4 sources: for $\log$\lledd $\approx -0.2$, ${\cal M} \approx 7.5$, $v_\mathrm{t} \approx  1.8 v_\mathrm{Kepl}$. If $v_\mathrm{Kepl} \approx 0.8 $HW(\hb), $v_\mathrm{t} \approx 1500$ \kms.  This value exceeds the peak velocity of the blue component in bin A3, and is in agreement with the ones measured in A4 and in SDSS J1508+4847.   Resonant line acceleration  might be needed if the gas has a  large  \nc\  ($\gtsim 10^{23}$ \cmq) or if the \mgii\ emitting gas is shielded by part of the continuum.  }

% {\sc cloudy} simulations  with a wind velocity field in a cylindrical geometry as well as with a static geometry   indicate that dense low-ionization gas  (\nh $\approx 10^{12} - 10^{13}$\cm3\ at inner radius, $\log U \ltsim -2$, \nc = 10$^{23}  - 10^{25}$\cmq) is a very strong \mgii\ emitter with intensity ratio \mgii/\hb $\gtsim$ 20. This effect alone might explain the difference in the \mgii\ and \hb\ line profiles.
  
\subsection{Alternate interpretations}  

{ Both \mgii\ and \hb\ line profile widths are probably modified by the viewing angle of the outflow/jet axis. Evidence exists that the line width of \hb\ is affected by  line-of-sight
orientation of the jet axis in
radio-loud sources
\citep{willsbrowne86,willsbrotherton95,rokakietal03,sulenticetal03,zamfiretal08}.
An effect on FWHM of a factor $\approx$2  is likely between core and
lobe-dominated sources. Orientation effects
are also expected for radio-quiet quasars
\citep[e.g.][]{jarvismclure06,punslyzhang10}.  Extreme and variable soft
X-ray emission from some narrow line Seyfert 1s (Pop. A) sources has been interpreted as a
signature of pole-on
orientation \citep[][and references therein]{sulenticetal00a}. Recent work confirms a
dependence on orientation for \hb\ in radio-loud sources and further
suggests a less-strong dependence
for \mgii\  \citep{runnoeetal12}. Following this line of reasoning the
occurrence of blueshifts in bin A3/A4 might involve
sources viewed at a favorable line-of-sight orientation. It is not clear
whether the results of \citet{runnoeetal12}
on different orientation sensitivity can be extended to Pop. A where
radio-loud sources are almost absent
in bins A2, A3 and A4. Even if the  FWHM(\hb) change could be explained on the basis of an orientation effect,  several line intensity ratios change
very strongly going from A1 to A4
\citep[e.g.,][]{willsetal99,aokiyoshida99,sulenticetal00a,bachevetal04,baldwinetal04,negreteetal12,shinetal12}:
\rfe\ by definition, but also \ciii/\siiii, \aliii/\siiii,
\siiv+\oiv/\civ. Emission line equivalent widths change as
well.  {For instance, the EW of \hb\ BC shows a  decrease from A1 to A4 by a factor $\approx$2 (Tab. \ref{tab:specfit}), as found previously \citep[e.g.,][]{sulenticetal00a}.  The large difference in EW persists also if the flux of the blueshifted component is included. }  Line intensity ratios and line equivalent widths are most likely
sensitive to density, ionization state and chemical composition of the gas
along with ionizing continuum shape (ultimately thought to be governed by Eddington ratio). In this respect we note that the decrease of $W$(\mgii) from $\approx$110 to $\approx$ 70 \AA\ is also consistent with a study showing an anticorrelation between $W$(\mgii) and \lledd\ \citep{dongetal09}.   It is unclear how
orientation might drive such changes. We are probably dealing with a
restricted range of \lledd\ in each spectral bin ``convolved'' with the
effect of orientation \citep{marzianietal01}. If a (rare) pole-on
orientation favors the observation of large shifts
then  SDSS J1508+4847 might be an example of a pole-on source.}
   
% In Pop. B sources  among lobe dominated a redshifted very-broad component is less prominent in \hb\ than in \mgii    
   
\section{Conclusion}

\mgii\ should be used as a virial estimator with caution { in high Eddington ratio sources}. Under the
simplest assumptions virial motion implies an unshifted and symmetric profile. A systematic line profile
blueshift can be interpreted as the signature of emission from radiatively acceleration of gas motion therefore invalidating the virial broadening assumption for \mgii\ in 20\%\ of Pop. A sources (10\%\ of all
quasars). { Conversely further work
has shown that the majority of quasars  show unshifted \mgii\ profiles
that are more symmetric than \hb\
(Sulentic et al. 2012 in preparation, and references therein). The width of \mgii\ is probably a
suitable virial broadening estimator
for those sources.}

%{ Almost constant \feiiuv/\mgii. Is it the same also for B sources?}

\acknowledgements PM acknowledges Junta de Andaluc\'{\i}a, through
grant TIC-114 and the Excellence Project P08-TIC-3531, and  the Spanish Ministry for Science and Innovation through grants AYA2010-15169 for supporting a sabbatical stay at IAA-CSIC. I.~P. - F. acknowledges the postdoctoral fellowship grants 145727
and 170304 from CONACyT Mexico. The authors wish to thank an anonymous referee whose suggestions helped them to critically reanalyze some of their results.  Funding for the SDSS and SDSS-II has been provided by the Alfred P. Sloan Foundation, the Participating Institutions, the National Science Foundation, the U.S. Department of Energy, the National Aeronautics and Space Administration, the Japanese Monbukagakusho, the Max Planck Society, and the Higher Education Funding Council for England. The SDSS Web Site is \url{http://www.sdss.org/}. Full acknowledgement of SDSS is given at \url{http://www.sdss.org/collaboration/credits.html}. 
\clearpage

\hoffset=-1.75cm
\begin{table*}
\begin{center}
\caption{Derived Quantities from the \hb\ and \mgii\ Profile   Multicomponent Analysis  \label{tab:specfit}}
\setlength{\tabcolsep}{2pt}
\begin{tabular}{lccccccccccccccccc}\hline\hline
\multicolumn{1}{c}{Sp. T.} &  \multicolumn{7}{c}{\hb} && \multicolumn{7}{c}{\mgii}\\ \cline{2-8} \cline{10-16}
& \multicolumn{3}{c}{BC} && \multicolumn{3}{c}{blue} && \multicolumn{3}{c}{BC} && \multicolumn{3}{c}{blue}\\ \cline{2-4} \cline{6-8}\cline{10-12}\cline{14-16}
 &\multicolumn{1}{c}{Int.$^\mathrm{a}$}& \multicolumn{1}{c}{Shift$^\mathrm{b}$} &  
\multicolumn{1}{c}{FWHM$^\mathrm{b}$} &&\multicolumn{1}{c}{Int.$^\mathrm{a}$}& \multicolumn{1}{c}{Shift$^\mathrm{b}$} &  
\multicolumn{1}{c}{FWHM$^\mathrm{b}$} &  &\multicolumn{1}{c}{Int.$^\mathrm{c}$} & 
\multicolumn{1}{c}{Shift$^\mathrm{b}$} & \multicolumn{1}{c}{FWHM$^\mathrm{b}$}  &&\multicolumn{1}{c}{Int.$^\mathrm{c}$} &\multicolumn{1}{c}{Shift$^\mathrm{b}$} &  
\multicolumn{1}{c}{FWHM$^\mathrm{b}$}\\ 
 \hline
%\multicolumn{14}{c}{{\em Pop. A Full Sample}} \\
A1	&	96	&	20$\pm$40		&	3180	&&	0.0	&	n.a.	&	n.a.	&	&	111	&	35$\pm$50	&	2710	&&	0.0	&	n.a.	&	n.a.	\\
A2	&	85	&	-20$\pm$40		&	2900	 &&	0.0	&	n.a.	&	n.a.	&	&	73	&	-70$\pm$50	&	2320	&&	0.0	&	n.a.	&	n.a.	\\
A3	&	45	&	45$\pm$40		&	2190	&&	12	&	-1240$\pm$230	&	4250	&	&	65	&	-150$\pm$50	&	2240	&&	0.0	&	n.a.	&	n.a.	\\
A3b	&	54	&	45$\pm$40	&	2190 	&&	6.5	&	-1420$\pm$510	&	4100 & 	&	54	&	0$^\mathrm{d}$	&	1750$^\mathrm{e}$ &&	5.6	&	-880$\pm$170	&	3100	\\
A4	&	28	&	70$\pm$50		&	1980	&&	16	&	-1240$\pm$220	&	4870	&	&	68	&	-265$\pm$90	&	2650	&&	0.0	&	n.a.	&	n.a.	\\
A4b	&	29	&	70$\pm$50		&	1940	&&	13	&	-1530$\pm$200	&	4460	&	&	46	&	0$^\mathrm{d}$	&	1585$^\mathrm{e}$	&&	10 	&	-1010$\pm$130	&	3300	\\														
J1508+48	&	32	&	-5$^\mathrm{d}$		&	2300	&& 	5.0	&	-1540$\pm$100 	&	4000 &	&	37	&	0$^\mathrm{d}$	&	2190	&&	20	&	-1490$\pm$160	&	3500	\\
\hline
\end{tabular}                   							
\begin{list}{}{}
\item[$^\mathrm{a}$]{Line intensity normalized to the continuum at 5050 \AA. The value  roughly corresponds to the rest-frame equivalent width in \AA. For J1508+48 values are in units of 10$^{-15}$\ergss\ \cmq\ \AA$^{-1}$. }
\item[$^\mathrm{b}$]{In units of \kms.}
\item[$^\mathrm{c}$]{Line intensity normalized to the continuum at 5050 \AA. The value can be used  an estimate of the \mgii/\hb\ intensity ratio. For J1508+48 values are in units of 10$^{-15}$ \ergss\ \cmq\ \AA$^{-1}$. }
\item[$^\mathrm{d}$]{Imposed to be consistent with rest frame.}
\item[$^\mathrm{e}$]{FWHM(\mgii) = 0.8 FWHM(\hb), for broad component.}
\end{list}
\end{center}
\end{table*}

%\hoffset=-1cm
%\begin{deluxetable}{lccccccc}
%\tablecaption{Derived Quantities from the \hb\ and \mgii\ Profiles}
%\setlength{\tabcolsep}{1pt}
%\tablewidth{0cm}
%\tabletypesize{\small}
%\tablehead{\colhead{Sp. Type}  
 %&\colhead{$\log$ \lbol}& \colhead{$\log L$(Mg{\sc ii})}  & \colhead{$f_\mathrm{c}$}& \colhead{$\log r_\mathrm{em}$}  & \colhead{$v_\mathrm{vir}$\tablenotemark{a}}   & \colhead{$\log$ \mbh}   
% \\
%\colhead{Source}& [\ergss] & [\ergss] &  & [cm] & [\kms]& [\msol]}
%\startdata
%A3	&	46.16 &	         43.40	&		0.13	& 17.6	& 1100 &	8.39   \\
%A4   &       46.20 &             43.53       &               0.15 & 17.7  & 1000 & 8.34 \\  
%SDSS J1508+4847	& 46.35   &    43.20  &  0.08	&	17.8	& 1000 &  8.46\enddata
%\tablenotetext{a}{HWHM of the BC of a single component of \mgii. }
%\tablecomments{ }
%\label{tab:tab}
%\end{deluxetable}
%
%\clearpage

\clearpage
\bibliographystyle{apj}

\end{document}